\documentclass[12pt]{iopart}

\expandafter\let\csname equation*\endcsname\relax
\expandafter\let\csname endequation*\endcsname\relax
\usepackage{graphicx}
\usepackage{amsmath}
\usepackage{color}

\begin{document}
\title{The manipulation of ultracold atoms of high orbitals in optical lattices}

\author{Shengjie Jin$^1$, Xuzong Chen$^1$ and Xiaoji Zhou$^1$}

\address{$^1$ State Key Laboratory of Advanced Optical Communication System and Network, School of Electronics, Peking University, Beijing, China}
\ead{\mailto{xjzhou@pku.edu.cn}}

\vspace{10pt}

\begin{abstract}
Ultracold atoms in optical lattices are a powerful tool for quantum simulation, precise measurement, and quantum computation. A fundamental problem in applying this quantum system is how to manipulate the higher bands or orbitals in Bloch states effectively. Here we mainly review our methods for manipulating high orbital ultracold atoms in optical lattices with different configurations. Based on these methods, we construct the atom-orbital qubit under nonadiabatic holonomic quantum control and Ramsey interferometry with trapped motional quantum states. Then we review the observation of the novel quantum states and the study of the dynamical evolution of the high orbital atoms in optical lattices. The effective manipulation of the high orbitals provides strong support for applying ultracold atoms in the optical lattice in many fields.
\end{abstract}

\today

\newpage
\tableofcontents
\newpage

\section{Introduction}

Ultracold atoms confined in optical lattices are a powerful tool for quantum simulation\cite{Greiner2002,Bloch2012,RevModPhys.80.885,Goldman2016,RevModPhys.80.885,RevModPhys.86.153,doi:10.1126/science.aal3837}, precision measurement\cite{PhysRevLett.124.120403}, and quantum computation\cite{Shui2021}. An optical lattice is formed by the interference of laser beams, creating a spatially periodic potential for ultracold atoms. The periodic potential generates the Bloch bands and Bloch states corresponding to different orbitals. Different to the study of ground band of optical lattice, by effectively manipulating the orbital degrees of freedom of ultracold atoms in the optical lattice, novel quantum states are found\cite{2011_Wirth_NatPhys,Niu_PRL2018,Wang2021,Jin_PRL2021}, and new qubits\cite{Shui2021} and interferometers\cite{Hu2018} based on the atomic orbitals are realized. In applying these high orbital ultracold atoms in optical lattices, a fundamental problem is how to manipulate the orbitals. Unlike internal states of atoms, the effective manipulation of Bloch states in optical lattices is complex because of the lack of selection rules\cite{NJP_zhou}. Recently, methods to effectively manipulate Bloch states and high orbitals of optical lattices have been proposed, such as stimulated Raman transitions\cite{PhysRevLett.99.200405}, shortcut method\cite{NJP_zhou,LXX2011,PhysRevA.87.063638,Hu2015,Yang2016,PhysRevA.94.033624,Guo2021}, phase imprint\cite{doi:10.1126/sciadv.1500854}, moving lattices\cite{PhysRevA.72.053605}, and band swapping technique\cite{2011_Wirth_NatPhys,PhysRevLett.106.015302,Wang2021,Jin_PRL2021}.

In this paper, we review our practical methods for manipulating high orbital atoms in optical lattices. These methods contain the shortcut method\cite{NJP_zhou}, band swapping technique\cite{2011_Wirth_NatPhys,PhysRevLett.106.015302,Wang2021,Jin_PRL2021}, and the amplitude modulation method to manipulate atoms in optical lattices\cite{Niu2015}. The shortcut is a nonadiabatic coherent control composed of lattice pulse sequences\cite{NJP_zhou}. The band swapping technique refers to quickly switching the structure of deep and shallow composite lattice, realizing the inversion of Bloch bands, and pumping atoms to the target band\cite{2011_Wirth_NatPhys}. The manipulation of atoms in an optical lattice by amplitude modulation is another flexible way to control Bloch states coherently. In this method, the modulation with more than one frequency is applied to the lattice, couples the different orbitals, and realizes a large-momentum-transfer beam splitter\cite{Niu2015}. We can manipulate any Bloch state of optical lattices with different configurations based on these methods.
All the experiments in this paper are based on a Bose-Einstein-Condensate (BEC) of $^{87}$Rb prepared in a hybrid trap with the harmonic trap frequencies $(\omega_x,\omega_y,\omega_z) = 2\pi\times (28,55,60)$ Hz\cite{Guo2021}. The experiments are mainly carried out in one dimensional (1D) and two dimensional (2D) triangular optical lattices. For 1D lattice, atoms are confined in more than 50 discrete pancakes with each pancakes containing 2000 atoms on average\cite{Niu_PRL2018}. For triangular lattice, there are around 800 tubes with each tube containing 60 atoms on average\cite{Jin_PRL2021}.

The structure of this review is organized as follows. In section \ref{methods}, we introduce the Bloch bands of atoms in optical lattices and review three methods for manipulating high orbital atoms in optical lattices. Then, we review the related applications based on these methods in section \ref{application}, including the atom-orbital qubit under nonadiabatic holonomic quantum control, Ramsey interferometry with trapped motional quantum states of the optical lattice. In section \ref{sec:porbital}, we review the observations of exotic quantum states of p-orbital ultracold atoms. In section \ref{dynamical}, we analyze the dynamical character of high orbital atoms in optical lattices. Finally, section \ref{summary} is the summary of this paper.

\section{Manipulation of high orbital atoms in optical lattices}\label{methods}
\subsection{Bloch bands of atoms in optical lattice}
An optical lattice is formed by the interference of a set of laser beams with electric field amplitude $\vec{E}_i$, which provides a spatially periodic potential for atoms:
\begin{equation}
	V(\boldsymbol{r})=\alpha \sum_{i,j} \boldsymbol{E}_i \cdot \boldsymbol{E}_j \cos((\boldsymbol{k}_i-\boldsymbol{k}_j) \cdot \boldsymbol{r} + (\beta_i - \beta_j))
\label{eq:V_lattice}
\end{equation}
where $\boldsymbol{k}_i$ is the wave number and $\beta_i$ is the initial phase of laser beams $i$, and $\alpha$ is the coefficient related to detuning, atomic energy level, etc. For red detuned beams, $\alpha$ is negative. After neglecting the interactions between atoms, the Hamiltonian for single atom in optical lattice is
$\hat{H} = \frac{\hat{p}^2}{2m} + V(\boldsymbol{r})$.
According to the Bloch's theorem, the periodic potential can generate Bloch bands and Bloch states $\Psi_{n,\boldsymbol{q}}$, which can be expressed by 
\begin{equation}
	\Psi_{n,\boldsymbol{q}} (\boldsymbol{r})  =u_{n,\boldsymbol{q}}(\boldsymbol{r})e^{i\boldsymbol{q}\cdot\boldsymbol{r}},
\label{eq:Bloch_state}
\end{equation}
where $n=1,2,3...$ is the index of the Bloch bands, $\boldsymbol{q}$ is the quasi-momentum, and $u_{n,\boldsymbol{q}}$ is a periodic function. The bloch states $\Psi_{n,\boldsymbol{q}}$, or the eigenstates of $\hat{H}$ can be projected onto a series of momentum eigenstates:
\begin{equation}
	\Psi_{n,\boldsymbol{q}} = \sum_{\ell_1,\ell_2,\ell_3} c_{\ell_1,\ell_2,\ell_3}^{n,\boldsymbol{q}} |\hbar (\ell_1 \boldsymbol{b}_1+\ell_2 \boldsymbol{b}_2+\ell_3 \boldsymbol{b}_3) + \boldsymbol{q} \rangle
\label{eq:Momentum_state}
\end{equation}
where $ \boldsymbol{b}_i$ is the reciprocal lattice vector.

\begin{figure}[h!]
\begin{center}
\includegraphics[width=14cm]{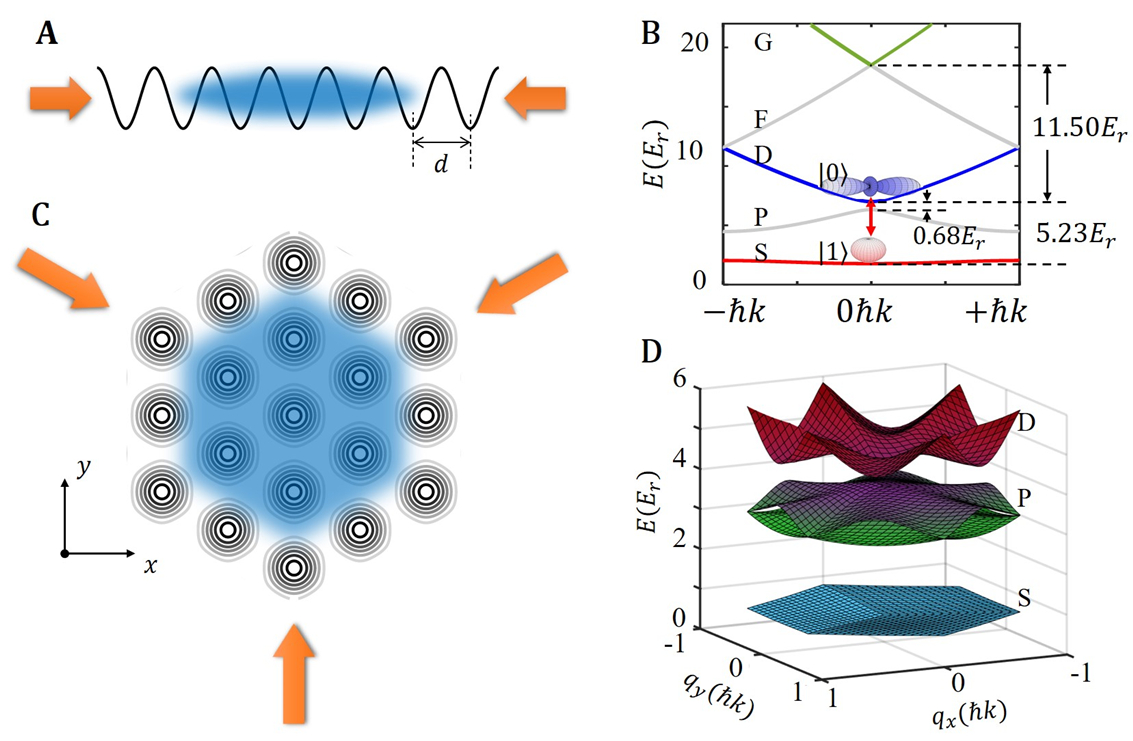}
\end{center}
\caption{ Schematic diagram and band structure of optical lattices. (\textbf{A}) is a 1D optical lattice with the lattice constant $d$. (\textbf{B}) The band-gap structures of 1D optical lattice for lattice depth $V=5E_r$ with laser wavelength $1064$ nm. (\textbf{C}) is a triangular optical lattice, with the corresponding band structure (\textbf{D}) for lattice depth $V=3E_r$.}\label{fig:lattice_bands}
\end{figure}

Figure \ref{fig:lattice_bands} shows two different configurations of optical lattices, 1D lattice and triangular lattice. As for the 1D lattice in Figure \ref{fig:lattice_bands}A, the lattice constant is $d=\lambda/2$ and the band spectrum is shown in figure \ref{fig:lattice_bands}B (when lattice depth $V=5E_r$, where $E_r=\hbar^2k^2/2m$ is the recoil energy and $k=2\pi/\lambda$). From the bottom to the top of the spectrum, the Bloch bands are S-, P-, D-, F-, and G-band, corresponding to $n=1,2,3,4,5$, respectively. In Figure \ref{fig:lattice_bands}C, the triangular lattice with lattice constant $2/3\lambda$, is constructed by three laser beams. 
The spectrum of this triangular lattice is different from that of the 1D lattice, where Bloch bands split into some different bands, as shown in figure \ref{fig:lattice_bands}D (when the lattice depth is $3E_r$).

Before introducing the manipulation of the atoms in Bloch bands, we review the methods used in our experiments to probe the Bloch states. The first method is the band mapping technique\cite{NJP_zhou,PhysRevLett.99.200405,PhysRevLett.94.080403,PhysRevLett.74.1542,PhysRevLett.87.160405}. In the experiment, if we switch off the lattice adiabatically within around 1 ms, the atoms in different bands can be mapped to different momentum components, measured after the time of flight (TOF). In addition, the value of quasi-momentum can be obtained from the position of atoms in corresponding Brillouin zones after the band mapping process and TOF. Figure \ref{fig:measurement}A shows the band mapping result of atoms in a triangular lattice. Atoms are mainly distributed in the fourth Brillouin zone, corresponding to the D-band.

The band mapping method is easy to implement in the experiment, and the atomic population of different Bloch bands can be obtained. However, in this mapping process, the phase information of the Bloch state is lost. Then we use a TOF quantum state tomography (TOFQST) method to extract full information about the Bloch states\cite{Shui2021}. For example, we use the TOFQST to detect the superposition of Bloch states in a 1D optical lattice with depth $V=5E_r$. In the experiment, we let the state evolve in the static lattice for a certain time $t_{evo}$. Then we turn off the lattice diabatically and measure the momentum distribution after 31 ms TOF, which is shown in figure \ref{fig:measurement}BC. By fitting the experimental data, we can obtain the full information of the states $\psi = \sqrt{0.493}\Psi_{1,0} + \sqrt{0.507}e^{i\cdot0.987\pi}  \Psi_{3,0}$.

\subsection{Shortcut to manipulating atoms in optical lattice}
\subsubsection{Introduction of the shortcut method}
\begin{figure}[h!]
	\begin{center}
		\includegraphics[width=18 cm]{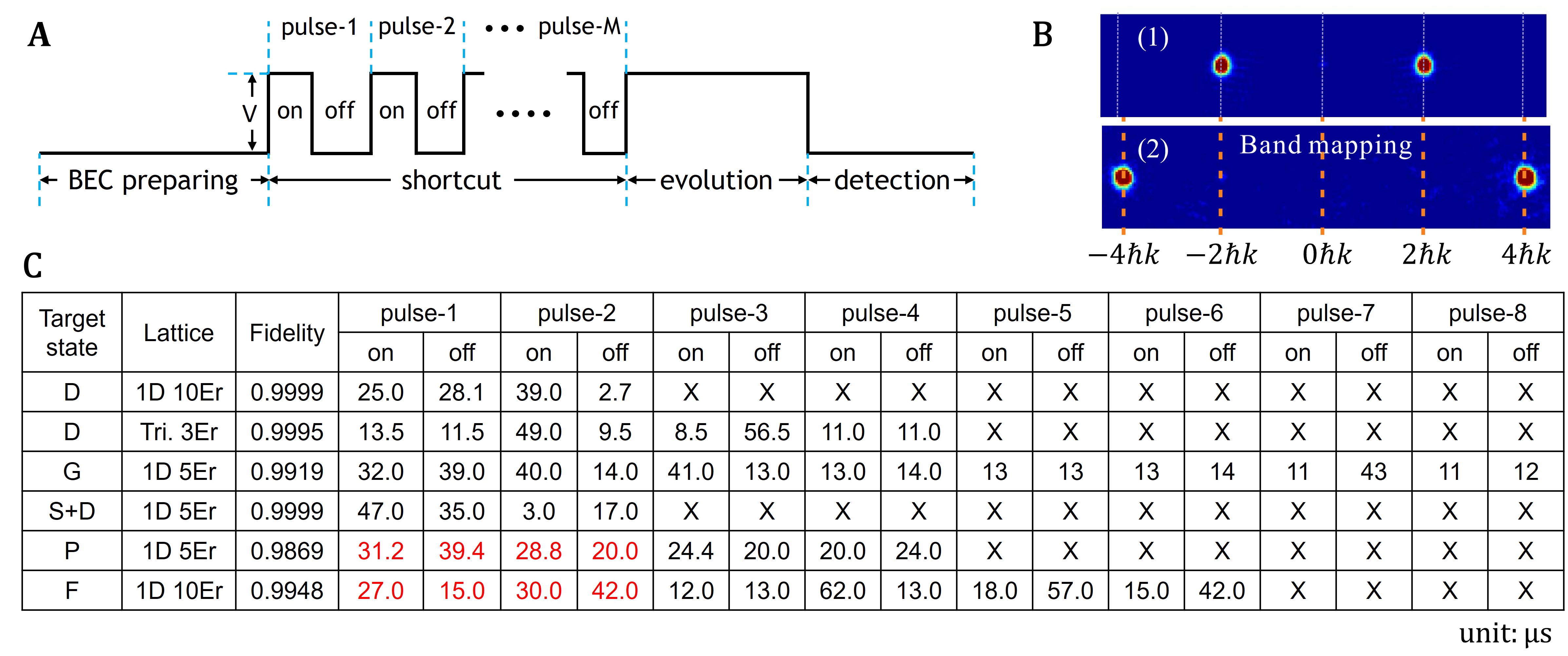}
	\end{center}
	\caption{ 
		(\textbf{A}) The timing diagram of the shortcut. (\textbf{B}) The band mapping result of D-band (top) and G-band (bottom) with zero quasi-momentum for 1D optical lattice. (\textbf{C}) The parameters of shortcut for different target states and optical lattice. The red time parameter is for the sequence of misplacement lattice for the P-band and F-band sequences. The quasi-momentum of all target states in this figure is zero. This symbol X indicates that there is no pulse.
	}\label{fig:shortcut_pulse}
\end{figure}

This section demonstrates a shortcut method for manipulating atoms in different Bloch bands\cite{NJP_zhou,LXX2011,PhysRevA.87.063638,Hu2015,Yang2016,PhysRevA.94.033624,Guo2021}. This method is characterized by short time and high fidelity, which can directly transfer ultracold atoms from the ground state in the harmonic trap to any Bloch state, and accurately manipulate atoms of different orbitals in optical lattices. The shortcut is composed of a series of optical lattice pulse sequences, which is shown in \ref{fig:shortcut_pulse}A. Each pulse pulse-i consists of two parts. First, the lattice is turned on for $t^{\rm{on}}_j$, and then the interval is $t^{\rm{off}}_j$. The time $\{t^{\rm{on}}_j,t^{\rm{off}}_j\}$ are optimized to achieve the goal of manipulating quantum states.

We consider a general situation for transferring an arbitrary initial state $|\psi_i\rangle$ to a target state $|\psi_t\rangle$, where the states $|\psi_i\rangle$ and $|\psi_t\rangle$ can be the Bloch eigenstates or the superposition states. This shortcut applied to the initial state can be expressed as an evolution operator $\hat{U}_s=\prod_{j=M}^{1}\hat{U}_j^{t^{\rm{on}}_j,t^{\rm{off}}_j}$. Here $\hat{U}_j$ represents evolution operator of the $j$th pulse, $\hat{U}_j=e^{-i(\hat{H}_j^{\rm{on}}t_j^{\rm{on}}+\hat{H}_j^{\rm{off}}t_j^{\rm{off}})}$, with $\hat{H}_j^{\rm{on}}= \frac{\hat{p}^2}{2m} + V(\boldsymbol{r})$ and $\hat{H}_j^{\rm{off}}= \frac{\hat{p}^2}{2m}$.
The final state after the shortcut is $|\psi_f\rangle=\hat{U}_s |\psi_i\rangle$. Then we define the fidelity $F$ of the manipulation
\begin{equation}
	F=|\langle\psi_f|\psi_t\rangle|=|\langle\psi_f|\prod_{j=M}^{1}\hat{U}_j^{t^{\rm{on}}_j,t^{\rm{off}}_j}|\psi_i\rangle|.
	\label{eq:Fidelity}
\end{equation}
By optimizing the pulse sequences $\{t^{\rm{on}}_j,t^{\rm{off}}_j\}$ to maximize $F$, the operation process with high fidelity for manipulating Bloch states or high orbital atoms in an optical lattice is obtained.

\begin{figure}[h!]
	\begin{center}
		\includegraphics[width=12cm]{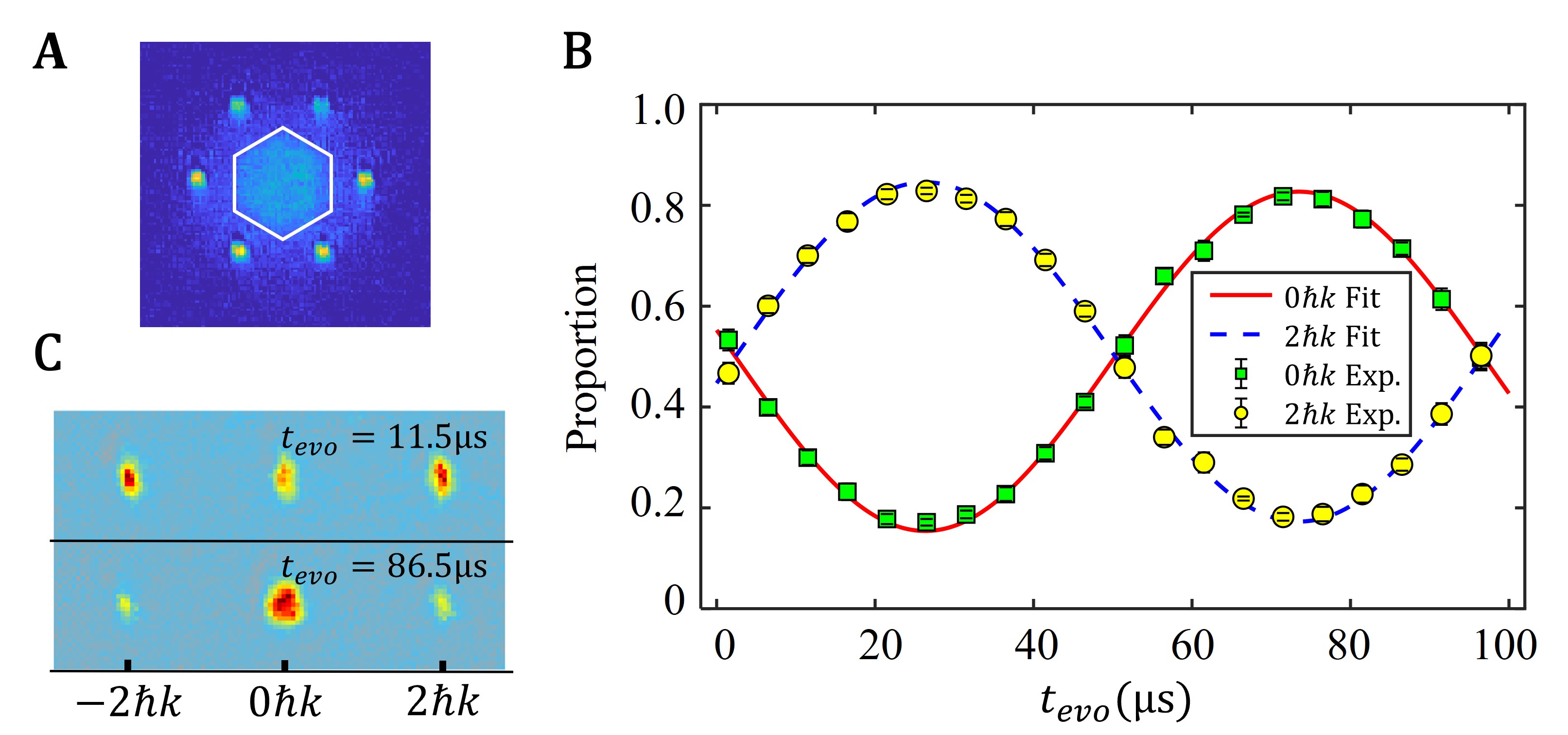}
	\end{center}
	\caption{ (\textbf{A}) Experimental result of atoms in D-band after band mapping process. The white hexagon is the boundary of the first Brillouin zone of the triangular lattice. The atoms are mainly distributed in the fourth Brillouin zone. (\textbf{B}) TOF quantum state tomography process. The initial state is a superposition Bloch state of the 1D optical lattice. After different evolution time $t_{evo}$, the atom number proportion at $0\hbar k$ and $\pm2\hbar k$ are shown by the green and yellow points. (\textbf{C}) The momentum distribution of atoms at $t_{evo}=11.5\rm{\mu s}$ and $86.5\rm{\mu s}$. The lattice depth is $V=5E_r$.}
	\label{fig:measurement}
\end{figure}

To verify the effectiveness of the shortcut, we demonstrate some experimental results in figure \ref{fig:measurement} and \ref{fig:shortcut_pulse}. In figure \ref{fig:shortcut_pulse}C, we list some pulse sequences for manipulating Bloch states, where the initial states are all the ground states of the harmonic trap, $|\boldsymbol{p}=0\rangle$ zero momentum state. For 1D optical lattice, we achieve to transfer atoms from zero momentum state to D-band, G-band, and the superposition states of S and D band with zero quasi-momentum, which corresponds to figure \ref{fig:shortcut_pulse}B1, B2, and figure \ref{fig:measurement}BC, and this theory fidelity is more than 99\%\cite{NJP_zhou}. For the D-band of the triangular optical lattice with depth $V=3E_r$, we give the shortcut sequence with 99.95\% fidelity in figure \ref{fig:shortcut_pulse}C, and the experimental band mapping results is shown in figure \ref{fig:measurement}A\cite{Guo2021}.

\subsubsection{Manipulating Bloch states with different symmetries}
The symmetry of Bloch states with different energy bands and quasi momentum is different. For example, in the 1D optical lattice, equation \ref{eq:Momentum_state} is rewritten as

\begin{equation}
	\Psi_{n,\boldsymbol{q}} = \sum_{\ell} c_{\ell}^{n,q} | \ell\cdot 2 \hbar k + q \rangle.
	\label{eq:1DMomentum_state}
\end{equation}

\begin{figure}[h!]
	\begin{center}
		\includegraphics[width=18 cm]{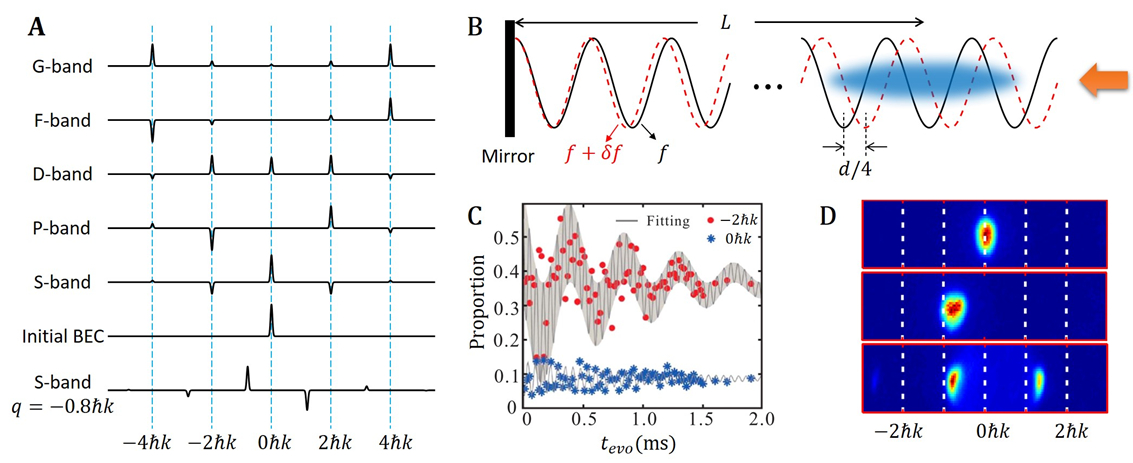}
	\end{center}
	\caption{ 
		(\textbf{A}) From the bottom to the top, the solid lines represent the momentum distributions of different states (S-band with $q=-0.8\hbar k$, initial BEC with $p=0$, S-, P-, D-, F-, and G-band with zero quasi-momentum). (\textbf{B}) Two 1D lattices with a $d/4$ position shift. (\textbf{C}) Population oscillations around momenta $-2\hbar k$(red points) and $0\hbar k$(blue stars). The solid lines are the fitting curves. (\textbf{D}) The experimental images: from top to bottom, the images are for the initial BEC with $p=0$, the BECs with $p_0=-0.8\hbar k$ after acceleration, and the final states $\Psi_{1,-0.8\hbar k}$.}\label{fig:symmetry}
\end{figure}

The parity of Bloch states on S-, D-, and G-bands with zero quasi-momentum is even, which is shown in figure \ref{fig:symmetry}A. On the contrary, the P- and F-bands with zero quasi-momentum are odd parity. Moreover, the symmetries of Bloch states with non-zero quasi-momentum are also different from that of zero quasi-momentum states. On the other hand, the symmetries of the 1D optical lattice and the initial states $|\boldsymbol{p}=0\rangle$ are even. Hence, we can easily prepare and manipulate even parity Bloch states, such as S-, D-, and G-bands with zero quasi-momentum. However, it is challenging to prepare states with different parity or change the symmetry. In order to expand the application scope of the shortcut, we propose two methods.

The first method is to use two misplacement lattices\cite{NJP_zhou,Hu2015,Niu_PRL2018}. In experiments, we realize this configuration through a reflective optical path and two laser beams with frequency difference $\delta f$, as shown in figure \ref{fig:symmetry}B. When the distance $L$ between BECs and mirror and laser frequency $f$ are fixed, we can realize the misplacement of ${d}/{4}$ of the two sets of standing wave optical lattices by adjusting the frequency difference $\delta f$. Because $\delta f \ll f$, the lattice constants of the two lattices can be regarded as the same. As an example, we transfer atoms from $|\boldsymbol{p}=0\rangle$ to P-band with zero quasi-momentum $\Psi_{2,0}$. We apply two sets of shortcut sequences in turn. The first one is of the lattice with frequency $f+\delta f$, and the second is of $f$. When the first sequence is switched to the second, each momentum state component $c_{\ell} | \ell\cdot 2 \hbar k  \rangle$ will be attached with a phase and becomes $e^{i\cdot\frac{\pi}{2}\ell}c_{\ell} | \ell\cdot 2 \hbar k  \rangle$. We design the first sequence to make $c_{\ell}=0$ for $\ell=2,4,6...$ and keep only $c_{\ell}$ for $\ell=1,3,4...$. So the result is that after the first sequence, the quantum state changes to odd parity, $c_{\ell}=-c_{-\ell}$ for all $\ell$. Finally, we use the second sequence to adjust this odd parity state to $\Psi_{2,0}$. In the experiment, we use TOFQST to detect the final state in the 1D lattice with depth $V=5E_r$, shown in figure \ref{fig:symmetry}C, and the experimental fidelity is more than 90\%. The parameters of these sequences are shown in the figure \ref{fig:shortcut_pulse}C\cite{Hu2015}. Similarly, we can also use this scheme to load atoms into the F-band\cite{NJP_zhou}, and the sequence is shown in figure \ref{fig:shortcut_pulse}C.

The second method is to change the symmetry of the initial state. For example, we demonstrate how to transfer atoms from initial BEC to non-zero quasi-momentum Bloch state in S-band. As shown in figure \ref{fig:symmetry}D, the atoms with $p=0$ is accelerated to obtain a momentum $p_0=-0.8\hbar k$. Immediately afterward, the designed shortcut sequence is used to transfer atoms into the S-band at $-0.8\hbar k$ quasi-momentum. The bottom figure of \ref{fig:symmetry}D is the momentum distribution of the final state in the experiment\cite{NJP_zhou}.

Hence, our shortcut method can be applied to manipulate arbitrary Bloch states in any Bloch band within a very short time and with high fidelity. Moreover, This shortcut can be applied to optical lattices with different configurations.

\subsubsection{Atomic momentum patterns with narrower intervals}
\begin{figure}[h!]
	\begin{center}
		\includegraphics[width=16 cm]{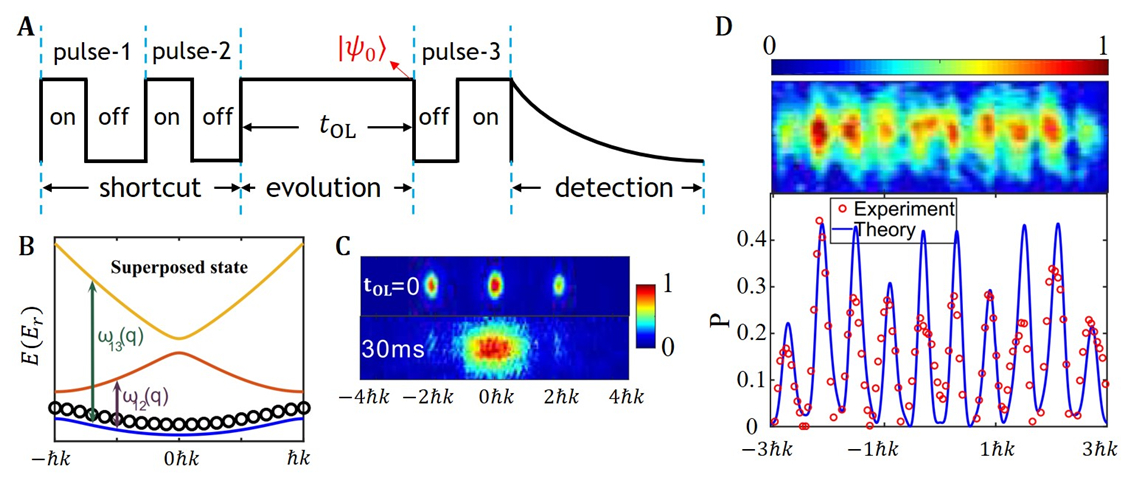}
	\end{center}
	\caption{ 
		(\textbf{A}) Time sequence: after the first two pulses and the 30 ms holding time in the optical lattice, the state becomes the superposition of the Bloch states in S-band with quasi-momentum taking the values throughout the first Brillouin zone, denoted as $|\psi(0)\rangle$. (\textbf{B}) The superposition of Bloch states of S-band spreading the whole S-band. (\textbf{C}) The comparison of the experimental results between $t_{OL}=0$ ms and $t_{OL}=30$ ms. These results are obtained after the sequence in (\textbf{A}) but without pulse-3. (\textbf{D}) Patterns with ten main peaks for the experimental measurements (the red circles) and the theoretical curves (the solid blue line).}\label{fig:narrower}
\end{figure}

The previous method is mainly for the Bloch state with a certain single quasi-momentum, and the intervals of the momentum peaks are $2\hbar k$ (for 1D optical lattice). The manipulation of the entire Bloch band and the preparation of narrow momentum peak distribution will also appear in many applications, such as the atom interferometer\cite{Robert_de_Saint_Vincent_2010,Impens_2011,RevModPhys.81.1051}. However, the manipulation of this superposition state with different quasimomenta is seldom studied\cite{Yang2016}.
In this section, we expand the shortcut to manipulating the superposed Bloch states with different quasi-momenta. Figure \ref{fig:narrower}A shows the time sequence for the manipulating process of the superposed states. The first challenge is to prepare the superposition states $|\psi(0)\rangle$ with different quasi-momentum that spread the whole S-band, as shown in figure \ref{fig:narrower}B. In the experiment, we first prepare a superposed state $\frac{1}{\sqrt{2}}(\Psi_{1,0} +\Psi_{3,0})$ by shortcut and hold the lattice for a time $t_{OL}$. When $t_{OL}=30$ ms, the atoms are almost all in the first Brillouin zone of S-band, which is due to collisions during the holding time, as shown in figure \ref{fig:narrower}C\cite{Yang2016}.

The next challenge is how to design the pulse-3 in figure \ref{fig:narrower}A for atomic momentum patterns with narrower intervals. First, we analyze the action of the pulse on $|\psi(0)\rangle$. The initial state can be expressed as
\begin{equation}
	|\psi(0)\rangle=\frac{1}{\sqrt{N}}\sum_{q=-\hbar k}^{\hbar k}\Psi_{1,q}.
	\label{eq:narrower_initial}
\end{equation}

After the pulse-3 (with interal $t^{\rm{off}}_3$ and duration $t^{\rm{on}}_3$), the momentum distribution $P(0,q)$ is
\begin{equation}
	P(0,q)\approx C_1 + C_2\cos(W_{1,0}t^{\rm{off}}_3) + C_3\cos(W_{-1,0}t^{\rm{off}}_3) + C_4\cos(\omega_{1,2}t^{\rm{on}}_3)+C_5\cos(\omega_{1,3}t^{\rm{on}}_3),
	\label{eq:narrower_p}
\end{equation}
where $C_i$ is the corrresponding amplitudes from the numerical calculations, $W_{\ell,\ell'}=\hbar^2 [(2\ell k +q)^2-(2\ell' k +q)^2]/2m$ and $\omega_{n,n'}=E^{n,q}-E^{n',q}$ (the band gap between $n$ and $n'$ band at quasi-momentum $q$) corresponds to the energy difference between different momentum states and Bloch states, respectively. By designing the $t^{\rm{off}}_3$ and $t^{\rm{on}}_3$, we can get the momentum patterns $P(0,q)$ with narrower intervals. In figure \ref{fig:narrower}D, we use the sequence $t^{\rm{off}}_3=118\rm{\mu s}$ and $t^{\rm{on}}_3=19\rm{\mu s}$, and obtain ten main peaks with $0.6\hbar k$ interval for lattice depth $10E_r$\cite{Yang2016}.

\subsection{Band swapping tecnique for loading atoms into high bands}\label{sec:swapping}
The band swapping technique is another method for loading atoms into high Bloch bands, which can be used to study the characteristics of Bloch bands and orbits. This technique is first proposed in \cite{2011_Wirth_NatPhys} to load atoms into the P-band of a checkerboard square lattice and is also applicable to other configurations of optical lattices. The key to realizing this technology is constructing a controllable composite optical lattice, including deep and shallow wells. Compared with the shortcut method, the band swapping technique is more suitable for the study of ground state or metastable state. 

\begin{figure}[h!]
	\begin{center}
		\includegraphics[width=18 cm]{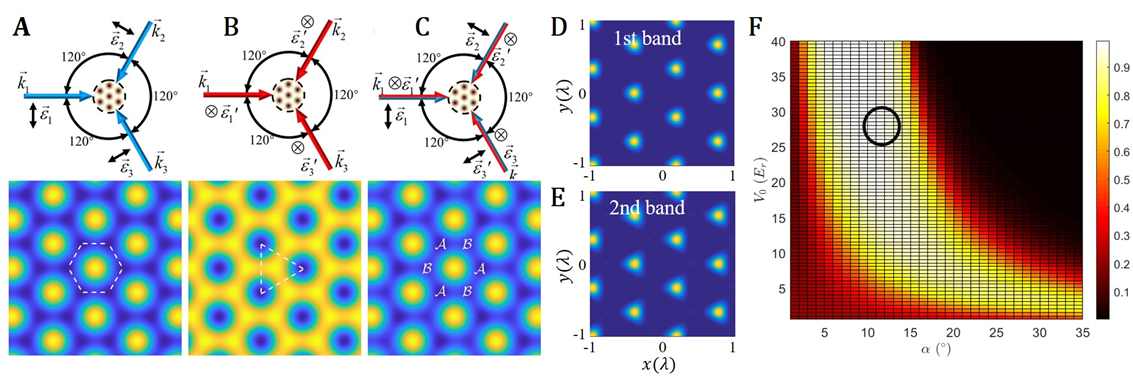}
	\end{center}
	\caption{The arrangement of the laser beams forms the honeycomb (\textbf{A}) and triangular (\textbf{B}) lattice. The two groups of sublattices form a composite hexagonal lattice (\textbf{C}), which consists $\mathcal{A}$ and $\mathcal{B}$ wells. (\textbf{D}) and (\textbf{E}) represent the wave function distributions in the real space of the ground band and the second band with zero quasi-momentum, respectively. (\textbf{F}) The wavefunction overlap with different lattice depth and light intensity ratios.}
	\label{fig:swapping}
\end{figure}

Here, we take a hexagonal lattice as an example to show the process of band swapping\cite{Jin_PRL2021}. We choose triangle lattice because it is more complex than square lattice, and it is impossible to separate variables in two directions directly. Three intersecting far-red-detuned laser beams form the hexagonal lattice in the $x-y$ plane with an enclosing angle of $120^{\rm{o}}$. Each beam is formed by combining two linearly polarized lights with polarization directions oriented in the $x-y$ plane (denoted as $\epsilon$ light) and along the $z$ axis (denoted as $\epsilon '$ light), respectively. The $\epsilon$ light form a honeycomb lattice as shown in figure \ref{fig:swapping}A. The $\epsilon '$ light form a triangular lattice as shown in figure \ref{fig:swapping}B. The superposition of the two groups of optical lattices is the configuration shown in figure \ref{fig:swapping}C, which consists of two wells with different depths (denoted as $\mathcal{A}$ and $\mathcal{B}$). The optical potential takes the form
\begin{equation}
	V(\boldsymbol{r})=-V_{\epsilon'}\sum_{i,j}\cos[(\boldsymbol{k}_i-\boldsymbol{k}_j)\cdot\boldsymbol{r}+\theta_j-\theta_i]+\frac{1}{2}V_{\epsilon}\sum_{i,j}\cos[(\boldsymbol{k}_i-\boldsymbol{k}_j)\cdot\boldsymbol{r}],
	\label{eq:hexagonal}
\end{equation}
where $\boldsymbol{k}_1=(\sqrt{3}\pi,-\pi)/\lambda$, $\boldsymbol{k}_2=(-\sqrt{3}\pi,-\pi)/\lambda$, $\boldsymbol{k}_3=(0,2\pi)/\lambda$, and $V_{\epsilon}$ ($V_{\epsilon'}$) is the depth of the honeycomb (triangular) lattice.
The depth difference between the two wells $\mathcal{A}$ and $\mathcal{B}$ can be adjusted by the $\epsilon$-to-$\epsilon'$ light intensity ratio (denoted as $\tan^2\alpha=V_{\epsilon'}/V_{\epsilon}$), and the relative phases $\theta_{1,2,3}$. First, we adiabatically load the BEC into the ground band with zero quasi-momentum of the optical lattice. The phase differences are initially set to be $\theta_{1,2,3}=(2\pi/3,4\pi/3,0)$, for which the $\mathcal{B}$ wells are deeper than the $\mathcal{A}$ wells. In real space, atoms mainly reside in the s orbitals of $\mathcal{B}$ wells as shown in figure \ref{fig:swapping}D. Then we switch the relative phases rapidly to $\theta_{1,2,3}=(4\pi/3,2\pi/3,0)$, making $\mathcal{A}$ wells much lower than $\mathcal{B}$. In this way, the atomic sample is directly projected onto the excited band. The key is to select appropriate parameters ($\epsilon '$ and $\epsilon$) to make the distribution of the wave function (figure \ref{fig:swapping}D) of the ground band before the switch consistent with that (figure \ref{fig:swapping}E) of the second band after the switch (at zero quasi-momentum). In our experiment, the range of parameter selection is in the black circle of the figure \ref{fig:swapping}F, where the total lattice depth is $30E_r$ and $\alpha=14^\circ$. As shown in figure \ref{fig:fig-nematicity}A, after this band swapping process, the atoms are pumped into the second band\cite{Jin_PRL2021}.

\subsection{Manipulation of atoms in optical lattice by amplitude modulation}\label{sec:modulation}
The modulation with more than one frequency component to optical lattices provides a flexible way to control quantum states coherently\cite{Niu2015}. In this section, we demonstrate bi-frequency modulations, which can be used to couple the S- and G-band of 1D optical lattice and realize a large-momentum-transfer beam splitter.

\begin{figure}[h!]
	\begin{center}
		\includegraphics[width=16 cm]{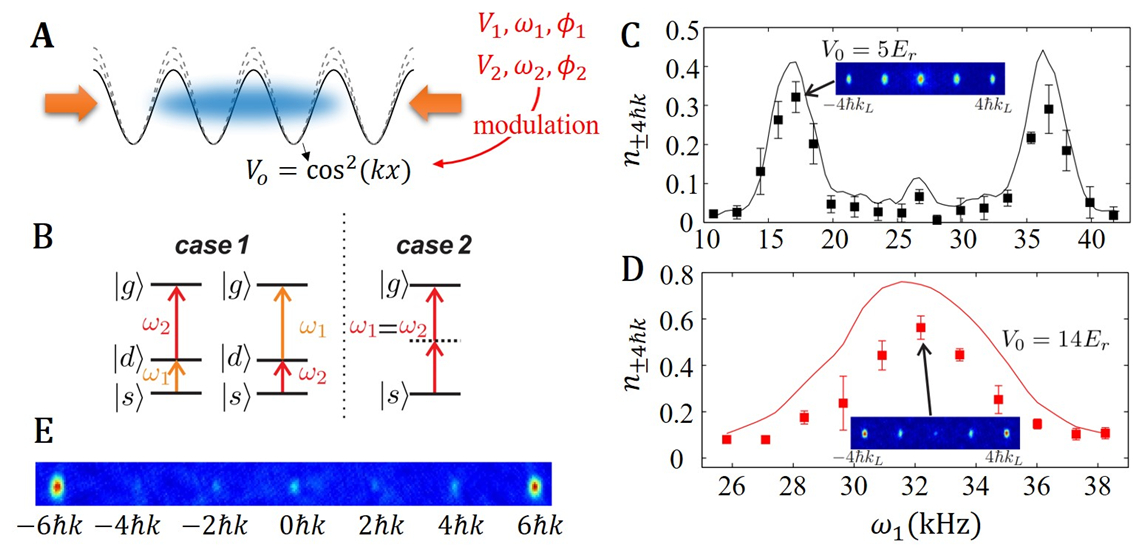}
	\end{center}
	\caption{(\textbf{A}) Schematic diagram of the polychromatic amplitude modulation lattice. (\textbf{B}) Two special cases in detecting the transfer population spectrum. In case 1, absorption of photons with $\omega_1$ or $\omega_2$ is resonant with D-band. In case 2, two frequencies are equal. Spectrum for the population on $\pm 4\hbar k$ states with increasing of modulation frequency $\omega_1$ for (\textbf{C}) $V_0=5E_r$ with $V_1=1.4E_r$, $V_2=1.6E_r$, modulation time $t=300\mu s$, and for (\textbf{D}) $V_0=14E_r$ with $V_1=V_2=2.5E_r$, and $t=150\mu s$. (\textbf{E}) A large-momentum-transfer beam splitter with a separation of $12\hbar k$.}
	\label{fig:modulation}
\end{figure}

For atoms in an amplitude modulated lattice system along the $x$ axis, as schematically shown in figure \ref{fig:modulation}A, the time-dependent Hamiltonian can be written as
\begin{equation}
	H(t)=\frac{p^2_x}{2m}+V_0\cos^2(kx)+\sum_i V_i\cos(\omega_i t+\phi_i)\cos^2(kx).
	\label{eq:modulation}
\end{equation}
The second term on the right hand represents optical lattice potential without modulation. The last term is the amplitude modulation with modulation amplitude $V_i$, the frequency $\omega_i$, and the phase $\phi_i$ of each frequency component\cite{Niu2015}.

According to the Floquet's theorem, a bi-frequency modulation induced two-photon process between S- and D-band is described by an effective Hamiltonian $H_{SG}$ as
\begin{equation}
	H_{SG}=\begin{pmatrix}
		E_S	&	e^{i\phi_1}V_1\Omega_{SD}	&	e^{i\phi_2}V_2\Omega_{SD}	&	0	&	0	&	0\\
		e^{-i\phi_1}V_1\Omega^*_{SD}	&	E_D-\hbar\omega_1	&	0	&	e^{i\phi_2}V_2\Omega_{DG}	&	e^{i\phi_1}V_1\Omega_{DG}	&	0\\
		e^{-i\phi_2}V_2\Omega^*_{SD}	&	0	&	E_D-\hbar\omega_2	&	e^{i\phi_1}V_1\Omega_{DG}	&	0	&	e^{i\phi_2}V_2\Omega_{DG}\\
		0	&	e^{-i\phi_2}V_2\Omega^*_{DG}	&	e^{-i\phi_1}V_1\Omega^*_{DG}	&	E_G-\hbar(\omega_1+\omega_2)	&	0	&	0\\
		0	&	e^{-i\phi_1}V_1\Omega^*_{DG}	&	0	&	0	&	E_G-2\hbar\omega_1	&	0\\
		0	&	0	&	e^{-i\phi_2}V_2\Omega^*_{DG}	&	0	&	0	&	E_G-2\hbar\omega_2
	\end{pmatrix},
	\label{eq:effect_H}
\end{equation}
where $\Omega_{nn'}=\langle \Psi_{n,0}|cos^2(kx)|\Psi_{n',0}\rangle$ with $\Psi_{n,0}$ the Bloch states on $n$ band at zero quasi-momentum.
We include six nearly degenerate states considering four main process in the excitation, which are $|E_S\rangle$, $|E_D-\hbar \omega_1\rangle$, $|E_D-\hbar \omega_2\rangle$, $|E_G-\hbar (\omega_1+\omega_2)\rangle$, $|E_G-2\hbar \omega_1\rangle$, and $|E_G-2\hbar \omega_2\rangle$. Using this basis a general state $(v_1,V_2,V_3,V_4,v_5,v_6)^T$ gives complex coefficient of the six dressed states. Population on S-band is $|v_1|^2$, while population on G-band is $|v_4e^{i(\omega_1+\omega_2)t}+v_5e^{2i\omega_1t}+v_6{2i\omega_2t}|^2$, given by coherent superposition of all G band states dressed with different Floquet photons.This effective model provides us a better understanding of the mutiphoton process. However, in the calculation more states associated with higher order processes could be included to get a more accurate result.

In the experiment, we keep the frequences $\omega_1$ and $\omega_2$ satisfying $\omega_1+\omega_2=\omega_{SG}$. According to the equation \ref{eq:effect_H}, there are two cases which would benefit the excitation process (shown in figure \ref{fig:modulation}B). 

\textbf{Case 1:} Resonant two-photon process. When $\omega_1=\omega_{SD}$ or $\omega_2=\omega_{SD}$, atoms are transferred from $\Psi_{1,0}$ to $\Psi_{5,0}$ with the assistance of D band as an intermediate band.

\textbf{Case 2:} Equal frequency two-photon process. When $\omega_1=\omega_2=\omega_{SG}/2$, two modulations with the same frequency can be added together, and the coupling strength of the process is doubled.

In the experiments, we sweep the frequency $\omega_1$ for different lattice depth $V=5E_r$ and $14E_r$, and measure the population on momentum states $\pm 4\hbar k$ that relect the transfer rate to G-band. For $V=5E_r$, in figure \ref{fig:modulation}C, two peaks appear at $\omega_1=\omega_{SD}$ and $\omega_2=\omega_{SD}$ follows \textbf{Case 1}. And the central peak at frequency $\omega_1=\omega_2=\omega_{SG}/2$ following \textbf{Case 2} is much lower than \textbf{Case 1}. For $V=14E_r$, only one peak is measured in figure \ref{fig:modulation}D, which means \textbf{Case 1} and \textbf{Case 2} are fulfilled simultaneously. Under this condition, the coupling is greatly enhanced. This bi-frequency modulation can also be applied to realize a large-momentum-transfer beam splitter, such as a distribution at $\pm 6\hbar k$. We choose the frequency $\omega_1=\omega_{SD}=\omega_{DG}$ and $\omega_2=\omega_{GI}$ (with I the 7th Bloch band). The experimental result is shown in figure \ref{fig:modulation}E\cite{Niu2015}.

This polychromatic amplitude modulation can be further improved to achieve more complex manipulation. We can control each period of the modulation waveform separately. Each period can be regarded as a pulse, which is shown in figure \ref{fig:Holonomic}A. The modulation amplitude $A_i$, phase $\phi_i$, and frequency $\omega_i$ are optimized to achieve more accurate Bloch states manipulation, such as holonomic quantum control\cite{Shui2021}, which will be discussed in detail next section.

In this section, we list three methods for manipulating ultracold atoms of high orbitals in optical lattices. The shortcut method is characterized by short time and high fidelity, which can directly transfer ultracold atoms from the ground state in the harmonic trap to any Bloch state, and accurately manipulate atoms of different orbitals in optical lattices. This method can be used to construct atomic orbital interferometers and qubits and to study the dynamic properties of high orbital atoms in optical lattices. The band swapping technique considers the interaction between atoms and the additional potential trap (such as harmonic trap) besides the optical lattice, which is more suitable for studying the ground and metastable states of the system. The amplitude modulation focuses on coupling different Bloch bands and can be used to realize quantum gates and the large-momentum-transfer beam splitter.

\section{Application of manipulating high orbital atoms in optical lattices}\label{application}

\subsection{Atom-orbital qubit under nonadiabatic holonomic quantum control}\label{sec:gate}
In section \ref{sec:modulation}, we mentioned that the amplitude modulation pulses could realize the holonomic quantum control. This section demonstrates an atom-orbital qubit by manipulating the s and d orbitals of BECs in the 1D optical lattice. Moreover, we achieve noise-resilient single-qubit gates by performing holonomic quantum control, allowing geometrical protection. The atom-orbital qubit and quantum control are based on the shortcut and amplitude modulation methods\cite{Shui2021}.  

As shown in figure \ref{fig:lattice_bands}A, the band gap between S and D band ($5.23E_r$) is much smaller than that between D and G band ($11.50E_r$) for optical lattice depth $V=5E_r$. With leakage to other bands neglected, the system corresponds to a two-level system, defining our atom-orbital qubit, $\Psi_{3,0}$ and $\Psi_{1,0}$ being orbital states identified as the qubit basis states $|0\rangle$ and $|1\rangle$. The 1D optical lattice potential is $V_p(x)=V_0\cos^2(kx)$, which is formed by 1064 nm laser beams. We use shortcut method to initialize the qubit to an arbitrary state, $|\psi\rangle=\cos\theta|0\rangle+\sin\theta e^{i\varphi}|1\rangle$. By TOFQST, we extract the fidelities of initial states $\{|0\rangle,|1\rangle,\frac{1}{\sqrt{2}}(|0\rangle+|1\rangle),  \frac{1}{\sqrt{2}}(|0\rangle-|1\rangle),\frac{1}{\sqrt{2}}(|0\rangle+i|1\rangle),\frac{1}{\sqrt{2}}(|0\rangle-i|1\rangle)\}$, and the averaged fidelity is 99.72(7)\%. The relaxation time and dephasing time are $4.5\pm0.1$ ms and $2.1\pm 0.1$ ms, respectively\cite{Shui2021}.

\begin{figure}[h!]
	\begin{center}
		\includegraphics[width=18 cm]{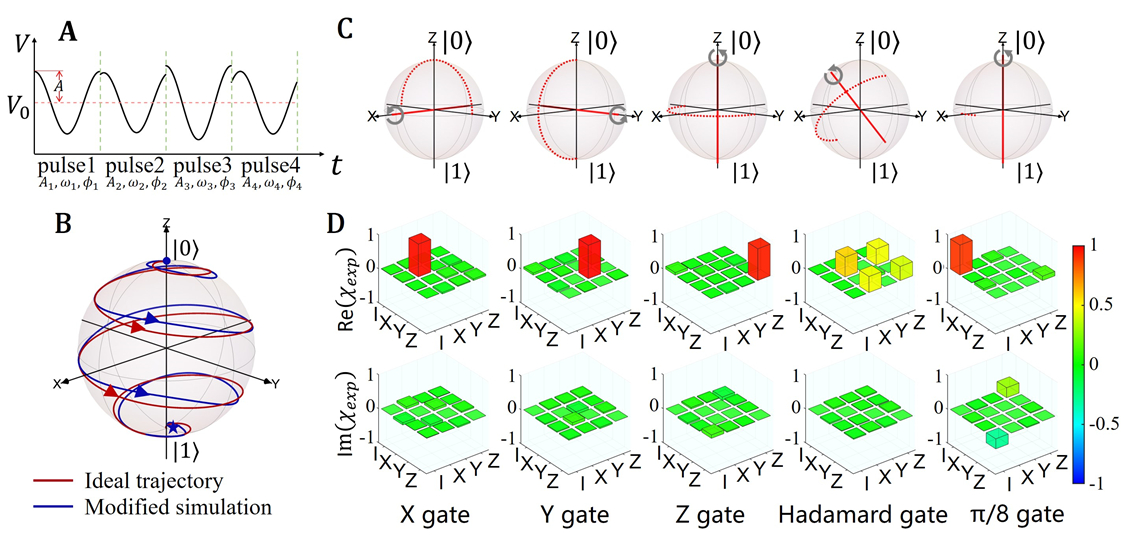}
	\end{center}
	\caption{(\textbf{A}) The pulse sequence of the holonomic gate. (\textbf{B}) Simulated time evolution of the $|0\rangle$ on the Bloch sphere under the holonomic X gate. (\textbf{C}) The schematic illustration of the quantum process. (\textbf{D}) Process matrices of the implemented holonomic X, Y, Z, Hadamard, and $\pi/8$ gates by quantum process tomography measurements.}
	\label{fig:Holonomic}
\end{figure}

The modulation pulses on potential takes form
\begin{equation}
	\Delta V(x,t)=A\sin(\omega t + \phi)V_p(x),
	\label{eq:gate_pulse}
\end{equation}
where amplitude $A$, phase $\phi$ and frequency $\omega$ programmable in our experiment. After a rotating wave approximation, we have a qubit control Hamiltonian $H(t)$
\begin{equation}
	H(t)=\frac{1}{2}\Delta \sigma_z + \frac{1}{2}\lambda [-\cos(\omega t + \phi)\sigma_y+\sin(\omega t+\phi)\sigma_x],
	\label{eq:gate_H}
\end{equation}
with $\Delta$ the gap between the S and D bands at zero quasi-momentum, and the 
\begin{equation}
	\lambda = A\int dxV_p(x)\Psi_{3,0}(x)\Psi_{1,0}(x),
	\label{eq:gate_lambda}
\end{equation}
We implement nonadiabatic holonomic orbital control base on dynamical invariant of the Hamiltonian in equation \ref{eq:gate_H}. To exploit the geometrical protection, the dynamical phase has to be canceled, which corresponds to
\begin{equation}
	\lambda^2+\Delta (\Delta-\omega)=0.
	\label{eq:gate_condition}
\end{equation}
Then, we calculate a control sequence of lattice modulation frequency, amplitude, and phase, denoted by $\Theta\equiv(\omega_i,A_i,\phi_i)$. After orbital leakage elimination, the gate fidelity is improved to above 98\% in the multiorbital numerical simulation. The simulated time evolution of the $|0\rangle$ on the Bloch sphere under the holonomic X gate is shown in figure \ref{fig:Holonomic}B. Besides, we also design the holonomic X, Y, Z, Hadamard, and $\pi/8$ gates. We further perform quantum process tomography to characterize the holonomic quantum gates, which is shown in figure \ref{fig:Holonomic}D. The measured quantum process fidelities are 98.47(9)\%, 98.35(11)\%, 97.81(13)\%, 98.53(8)\%, 98.63(15)\%, and 98.63(15)\%, for the X, Y, Z, Hadamard, and $\pi/8$ gates, respectively\cite{Shui2021}.
There are four main factors limiting fidelities:
a) orbital leakage: although we have greatly eliminated the band leakage, there are still a small number of atoms in high bands that affect the current fidelities.
b) De-phasing mechanism: the quasi-momentum broadening of BEC and the non-uniformity of light intensity of optical lattice will lead to the de-phasing effect, which causes the decrease of fidelity.
c) Atom interaction: As discussed in section \ref{D-collision}, atoms in the D band will be scattered to the ground band due to collision, which will affect the fidelities.
d) Error caused by measurement: It is mainly caused by the vibration of the imaging system during the absorption imaging process.
For a) and b), we can overcome them through super lattice or more complex lattice. We can construct a more perfect two-level so that the G-band is far away from the D-band. Further, by constructing flat bands, the dephasing effect brought by momentum broadening and non-uniformity of the light can be suppressed, so as to greatly improve the fidelity. 
If we want to further improve the fidelity, we need to consider the factors of c) and d). We can use Feshbach resonance technology to reduce the atomic interaction, and reduce the imaging error through the improvement of the mechanical structure of the imaging system.

\subsection{Ramsey interferometry with trapped motional quantum states of the optical lattice}\label{RamseyI}
Ramsey interferometry (RI) using internal electronic or nuclear states find wide applications in science and engineering\cite{PhysRevLett.67.177}. In this section, we review a new RI with the S- and D-bands in an optical lattice\cite{Hu2018}, similar to figure \ref{fig:lattice_bands}B (in this section, the laser wavelength of the optical lattice is 852 nm, and the lattice depth is $10E_r$). A key challenge for constructing this RI is to realize $\pi$-pulse and $\pi/2$-pulse analogous to those in conventional RIs. In section \ref{sec:gate}, we have achieved the arbitrary holonomic quantum control, which ensures the noise-resilient but increases the control time. However, we want to study the quantum many-body lattice dynamics by this RI, which requires the time duration of the $\pi$-pulse and $\pi/2$-pulse to be as short as possible. Hence, we use the shortcut method to design these pulse sequences.

\begin{figure}[h!]
	\begin{center}
		\includegraphics[width=18 cm]{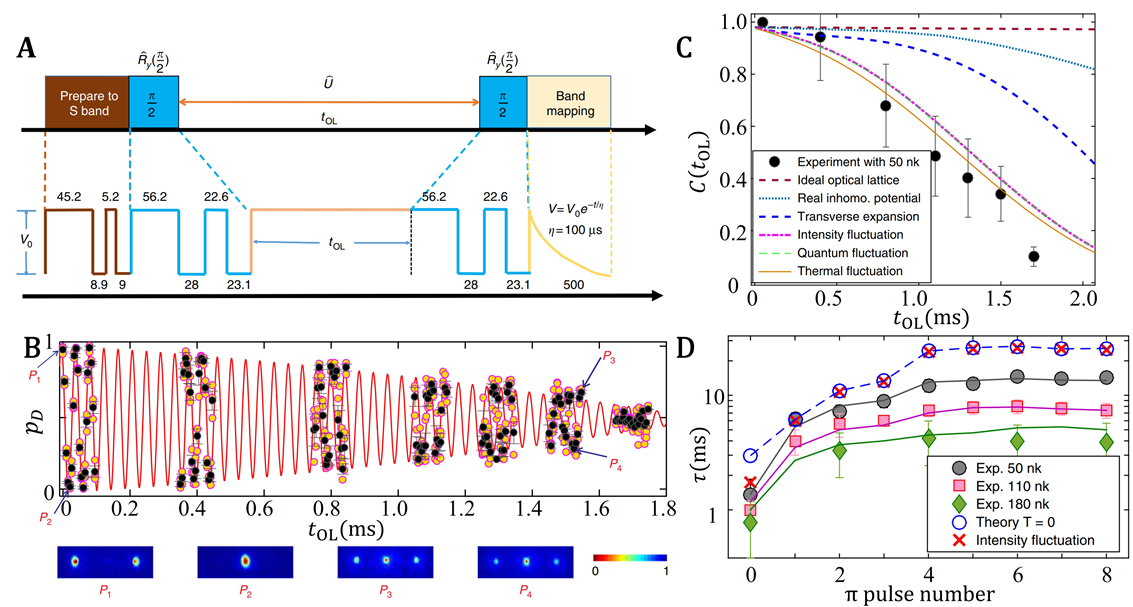}
	\end{center}
	\caption{(\textbf{A}) Time sequences for the Ramsey interferometry. The atoms are first loaded into the S-band, followed by the RI sequence $\pi/2$-pulse, holding time $t_{OL}$, and the second $\pi/2$-pulse. Finally, band mapping is used to detect the number of atoms in the different bands. The used sequences designed by the shortcut method are shown in the figures (unit: $\rm{\mu s}$). (\textbf{B}) The oscillation of the population of atoms in the D-band $p_D$. The images below show typical time of flight pictures after band mapping. (\textbf{C}) The experimental contrast (black points) and the theoretical calculation for $V_0=10E_r$. In theory, we begin with an ideal optical lattice and gradually add the effect of lattice inhomogeneity, transverse expansion of atom cloud, intensity fluctuation of laser amplitude, quantum fluctuation, and thermal fluctuation. (\textbf{D}) Coherence time $\tau$ vs. the number of applied $\pi$-pulse n with different temperatures.}
	\label{fig:Ramsey}
\end{figure}

The full time sequence for RI with two shortcut $\pi/2$-pulses $\hat{R}(\pi/2)$ we use in experiments is shown in figure \ref{fig:Ramsey}A. First the atoms in the harmonic trap are transferred into the S-band by shortcut, then the first pulse $\hat{R}(\pi/2)$ is applied to prepare an initial superposition state $\frac{1}{\sqrt{2}}(\Psi_{1,0} +\Psi_{3,0})$. After evolution in the optical lattice for time $t_{OL}$ and a second $\pi/2$-pulse, the final state can be expressed as $\Psi_f=a_S\Psi_{1,0}+a_D\Psi_{3,0}$. Then we apply band mapping to read out the final state, and obtain the population of atoms in S (D) band, denoted as $N_S$ ($N_D$). We define the signal of this RI as $p_D(t_{OL})=N_D/(N_S+N_D)$. For the ideal single-atom system, where the imperfection and decoherence can be neglected, the signal $p_D(t_{OL})=[1+\cos(\omega t_{OL}+\phi)]/2$, with $\omega$ corresponding to the energy difference between $\Psi_{1,0}$ and $\Psi_{3,0}$. However, when $t_{OL}$ gets longer, the oscillation amplitude decays, as shown in figure \ref{fig:Ramsey}B. The contrast $C(t_{OL})$ can be obtained by fitting the amplitude of oscillation $p_D(t_{OL})$ with
\begin{equation}
	p_D(t_{OL})=[1+C(t_{OL})\cos(\omega t_{OL}+\phi)]/2.
	\label{eq:contrast}
\end{equation}

In order to improve the performance of the RI, we now investigate the mechanisms that lead to RI signal attenuation. The decay of the contrast, shown by the black points in figure \ref{fig:Ramsey}C, mainly comes from de-phasing and de-coherence mechanisms. These mechanisms are caused by the imperfect design of the $\pi/2$ pulse, non-uniform potential distribution of the Gaussian beam in the radial direction, atom-atom interaction leading to transverse expansion, intensity fluctuation of the lattice, and thermal fluctuations of the system. The theoretical calculation with different mechanisms is shown in figure \ref{fig:Ramsey}C, which is consistent with the experimental results. The expansion leads to a significant reduction in contrast (blue dashed line in figure \ref{fig:Ramsey}C). In our experiment, the influence of the quantum fluctuation is not significant.

To further improve the contrast of the RI, we develop a matter-wave band echo technique. A $\pi$-pulse is designed, which swaps the atom population in the S- and D-band. The $\pi$-pulse is inserted into the center of the evolution process. After implementing one $\pi$-pulse, the intensity fluctuation can be well suppressed. After applying six echo pulses, the effects of non-uniform lattice potential and transverse expansion are eliminated. The coherence times $\tau$ (defined as the time when the contrast $C(t_{OL})$ drops to $1/e$) with different echo pulses are shown in figure \ref{fig:Ramsey}D. The coherent time is increased from 1.3 ms to 14.5 ms by the echo pulses\cite{Hu2018}.

\section{Exotic quantum states of p-orbital ultracold atoms in optical lattices}\label{sec:porbital}
\subsection{Observation of a dynamical sliding phase superfluid with P-Band bosons}
\begin{figure}[h!]
	\begin{center}
		\includegraphics[width=16 cm]{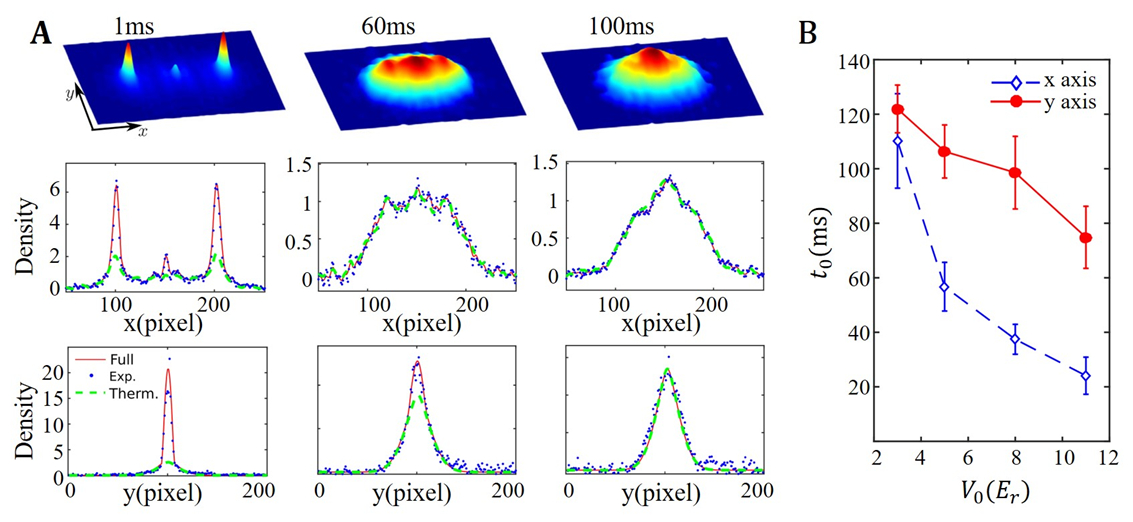}
	\end{center}
	\caption{(\textbf{A}) Momentum distributions measured at three different holding times $t_{evo}=1,60,100$ ms, along different directions (with $V_0=5E_r$). The corresponding distribution along the x-direction is given in the second row with blue dots. The solid red line gives a full fitting line, while the green dashed line gives the distribution of the thermal component. The third row shows the atomic distribution in the y-direction for images measured by probe 2. (\textbf{B}) The time $t_0$ for the atoms to lose coherence in lattice and pancake directions with different optical-lattice depths. The blue diamonds are for the x-direction (lattice) by probe 1 and the dotted points are for the y-direction (pancake). The optical lattice is the 1D lattice.}
	\label{fig:fig-sliding}
\end{figure}

The Sliding phase\cite{PhysRevLett.83.2745}, which is introduced to characterize intricate phase transitions in a wide range of the many-body system, appears under extreme conditions for thermal equilibrium systems or quantum ground states, causing a grievous challenge in experimental implementation\cite{PhysRevB.33.4767,PhysRevB.31.4516,FEIGELMAN1990177,PhysRevB.12.877,PhysRevLett.73.1384,PhysRevLett.59.1112,PhysRevLett.80.4345}. Here, we review the observation of a sliding phase superfluid in a dynamical system of ultracold atoms in the P-band\cite{Niu_PRL2018}. We load the atoms into P-band with zero quasi-momentum by the shortcut as shown in figure \ref{fig:symmetry}. The quantum system is driven to a far-out-of-equilibrium but a phase-coherent state. We hold the condensate in the P-band for a certain time $t_{evo}$. Then the TOF images are taken in two probe directions with probe light along the z-axis (denoted as probe-1) and along the x-axis (denoted as probe-2). The distribution in different probe directions is analyzed via a bimodal fitting, as shown in figure \ref{fig:fig-sliding}A. We extract the coherent fraction from the bimodal fitting so that the phase coherence of the dynamical many-body state can be inferred\cite{Niu_PRL2018}. 

From the time evolution, we identify three distinct dynamical regions. At an early time, the system has superfluid phase in all directions, as shown in figure \ref{fig:fig-sliding}A at $t_{evo}=1$ ms. In the second state, the phase coherence of the quantum gas survives partially. The bimodal fitting in figure \ref{fig:fig-sliding}A at $t_{evo}=60$ ms shows that there is a finite condensed component in the pancake directions, but no such component in the lattice direction, which is called the sliding phase. At the last state, the quantum gas has rethermalized with a complete loss of phase coherence. The bimodal fitting shows that all atoms are thermal in the complete absence of any condensed component. We define the lifetime $t_0$ for the atoms to lose coherence in lattice and pancake directions, which is shown in figure \ref{fig:fig-sliding}B for different lattice depth. We find that with the total atom number fixed in the experiment, a critical lattice depth appears beyond which the sliding phase superfluid starts to emerge. Moreover, we also verify that P-band is necessary to realize the dynamical sliding phase in our experiment, and the sliding phase is absent for cold atoms in the S-band at equilibrium\cite{Niu_PRL2018}.

\subsection{Observation of a Potts-Nematic superfluidity in a Hexagonal $sp^2$ optical lattice}
\begin{figure}[h!]
	\begin{center}
		\includegraphics[width=12 cm]{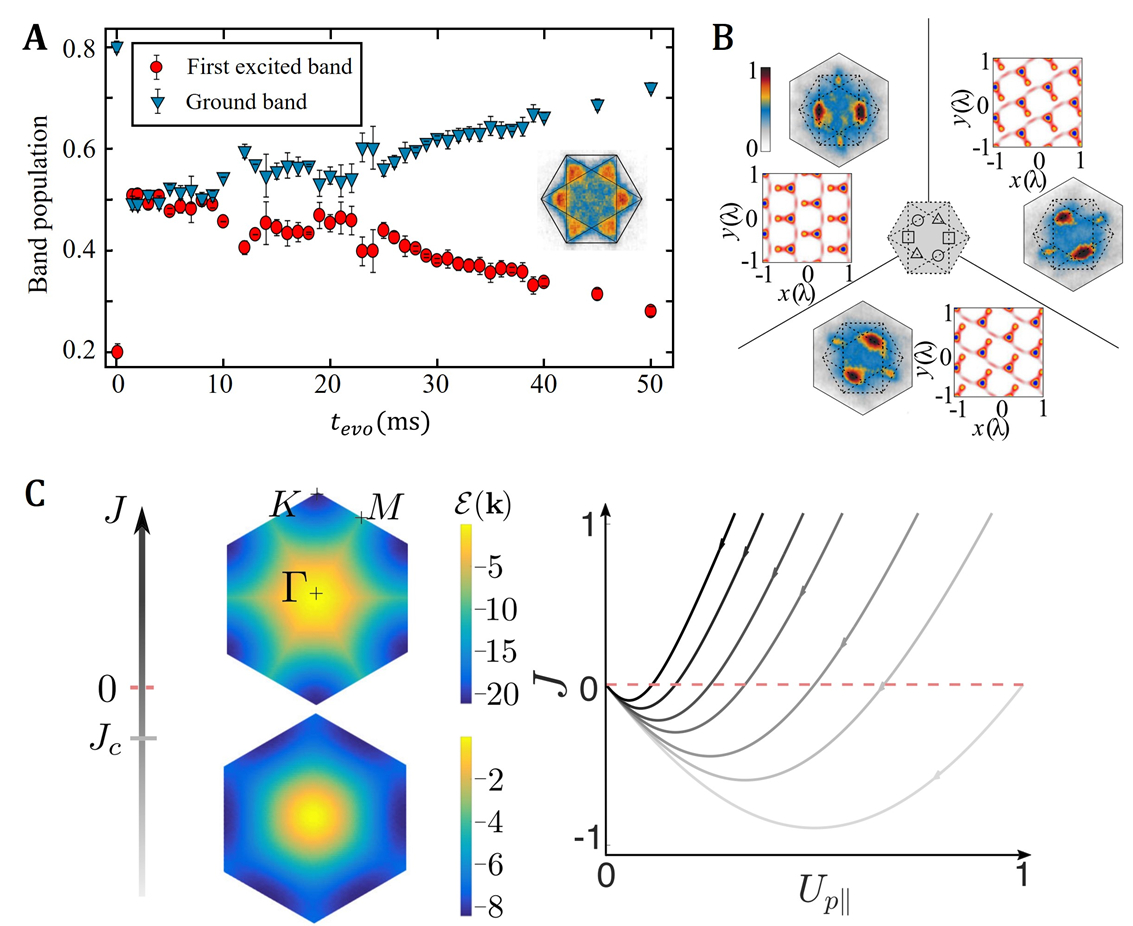}
	\end{center}
	\caption{(\textbf{A}) The measured time evolution of the atomic population in the ground and the second bands of the hexagonal optical lattice. \textbf{B} The averaged momentum distribution for three symmetries. \textbf{C} Theoretical quantum phase transitions varying the orbital Josephson coupling. The energy $\varepsilon(\boldsymbol{k})$ for a plane-wave condensate at a lattice momentum $\boldsymbol{k}$. The energy $\varepsilon(\boldsymbol{k})$ has minima at K (M) points for $J>J_c$ ($J<J_c$). The right figure is the sketch of the renormalization of the p-orbital couplings to low energy.}
	\label{fig:fig-nematicity}
\end{figure}

In this section, we review the observation of a Potts-nematic quantum state in a system of cold atoms loaded into the second band of a hexagonal optical lattice\cite{Jin_PRL2021}. We use the band swapping method to load the atoms into the second band, as shown in section \ref{sec:swapping}. After the atoms are transferred into the band maximum of the second band, the phase coherence in the state will immediately disappear. After a few milliseconds, the phase coherence reemerges, and the quantum state spontaneously chooses one orientation, giving rise to a three-state Potts nematicity. We divide the experimental TOF images into three classes and take the average within each class\cite{Jin_PRL2021}. The post-classification averaged results are shown in figure \ref{fig:fig-nematicity}B.

To gain insight into the mechanism supporting the Potts-nematic order in the $sp^2$-orbital hybridized band, we provide a mean-field theory analysis assuming a plane-wave condensate. The interaction can be expressed as
\begin{equation}
	H_{int}=\frac{1}{2}\sum_{\boldsymbol{r}\in\mathcal{B}}U_s \hat{s}_{\boldsymbol{r}}^\dagger \hat{s}_{\boldsymbol{r}}^\dagger \hat{s}_{\boldsymbol{r}} \hat{s}_{\boldsymbol{r}}+
	\sum_{\boldsymbol{r}\in\mathcal{A}}\{J[\hat{p}_{x,\boldsymbol{r}}^\dagger \hat{p}_{x,\boldsymbol{r}}^\dagger \hat{p}_{y,\boldsymbol{r}} \hat{p}_{y,\boldsymbol{r}}+H.c.] \}+
	\frac{1}{2}\sum_{\boldsymbol{r}\in\mathcal{A}}\sum_{\alpha,\beta\in\{x,y\}}U_{p,\alpha \beta}\hat{p}_{\alpha,\boldsymbol{r}}^\dagger \hat{p}_{\beta,\boldsymbol{r}}^\dagger \hat{p}_{\beta,\boldsymbol{r}} \hat{p}_{\alpha,\boldsymbol{r}}
	\label{eq:pinter}
\end{equation}
where $\hat{s}$ and $\hat{p}$ represent quantum mechanical annihilation operators for s and p orbitals, and the p-orbital couplings are constrained by $U_{p,xx}=U_{p,yy}\equiv U_{p_\parallel}$, $U_{p,xy}=U_{p,yx}\equiv U_{p_\perp}$, $J=U_({p_\parallel}-U_{p_\perp})/2$. Taking a trial condensate wave function with $\langle s_{\boldsymbol{r}}\rangle=\phi_s e^{i\boldsymbol{k}\cdot\boldsymbol{r}}$, $\langle p_{x,y,\boldsymbol{r}}\rangle=\phi_{x,y} e^{i\boldsymbol{k}\cdot\boldsymbol{r}}$, with $\phi_s$, $\phi_{x,y}$ the variational parameters. For each lattice momentum $\boldsymbol{k}$, we minimize the energy by varying $\phi_{s,x,y}$, and the resultant energy is denoted as $\varepsilon(\boldsymbol{k})$ and shown in figure \ref{fig:fig-nematicity}C. With the orbital Josephson coupling $J>0$, both the kinetic and interaction energies favor a condensate at K points which breaks the time-reversal symmetry but respects the rotation symmetry. With the Josephson coupling $J<0$, minimizing the kinetic and interaction energies meets frustration, as interaction favors p-orbital polarization. Figure \ref{fig:fig-nematicity}C shows the possible value range of J considering renormalization effects\cite{Jin_PRL2021}.

\section{Dynamical evolution for atoms in high bands of optical lattices}\label{dynamical}
\subsection{The scattering channels induced by two-body collision of D-band atoms in optical lattices}\label{D-collision}
\begin{figure}[h!]
	\begin{center}
		\includegraphics[width=18 cm]{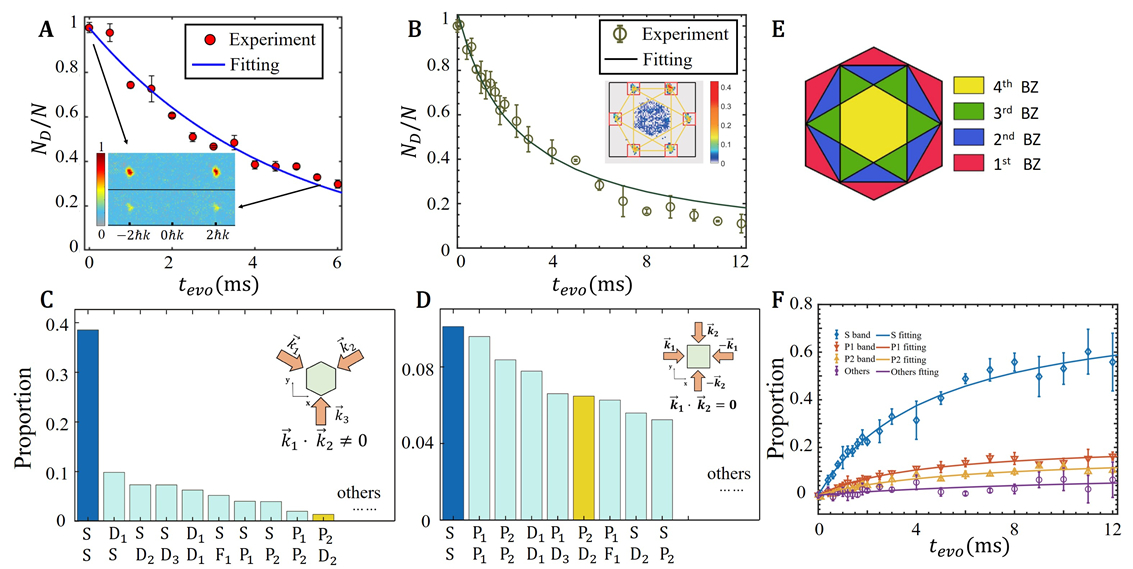}
	\end{center}
	\caption{(\textbf{A}) The proportion of D-band over the evolution time in 1D lattice. (\textbf{B}) The proportion of the first D-band (D1-band) in the triangular lattice. The proportions of several main scattering channels are shown in (\textbf{C}) (triangular lattice) and (\textbf{D}) (square lattice). The inserts show the diagram of a square optical lattice and triangular optical lattice. (\textbf{E}) Schematic diagram of the first four Brillouin zones of a triangular optical lattice. Yellow, green, blue, and red areas represent $1st, 2nd, 3rd, 4th$ BZ, respectively. (\textbf{F}) The atomic proportion in S (P1, P2 band, and others), of which the solid lines with the same color fit lines.}
	\label{fig:scatter}
\end{figure}

The mechanism of atomic collisions in excited bands plays an essential role in the atomic dynamics in high bands of optical lattices and the simulation of condensed matter physics\cite{Guo2021}. Atoms distributed in an excited band of an optical lattice can collide and decay to other bands through different scattering channels\cite{PhysRevA.87.063638,Guo2021}. The decay rate and scattering channels of optical lattices with different configurations are different. Here, we first compare the lifetime of atoms in the D-band for one-dimensional lattice and triangular lattice. In experiments, we utilize the shortcut method to load BECs to the D (D1) band of the 1D optical lattice (triangular optical lattice). Then the BECs in the optical lattice evolve for a certain time $t_{evo}$. Finally, we apply band mapping to measure the proportion $p_D(t_{OL})$ of atoms in the D-band, which is shown in figure \ref{fig:scatter}A for 1D lattice and \ref{fig:scatter}B for triangular lattice. We define the lifetime $\tau$ of atoms in the D-band as the proportion $p_D$ reduces to $1/e$. The lifetime for the triangular lattice is 5.0 ms, which is much longer than that of the 1D lattice, 2.1 ms. The difference in collisional scattering cross-section leads to the difference of a lifetime. Next, we will carefully analyze the scattering cross-section and the scattering channels\cite{Guo2021}.

We take triangular and square lattices as examples to study the difference in the scattering process. We use the scattering theory to calculate the cross-section of each scattering channel in those two types of lattices.
Two-body collisional scattering cross section for two atoms initially at the $\Gamma$ point ($(q_x,q_y)$=(0,0)) of D1 band jumping to band $n_1$ and $n_2$ can be written as:
\begin{equation}
	\label{sigma_s}\sigma(n_1 ,n_2)=\frac{4\pi m \hbar}{v_{a}}\int d\boldsymbol{q} \times |-2\pi i \frac{4\pi a_s}{m} \zeta_{n_1 ,n_2}(0,0;\boldsymbol{q},-\boldsymbol{q})|^2, 
\end{equation}
where $v_{a}$ is the atomic velocity and $a_s$ is atomic $s$-wave scattering length. And the overlapping integral of eigenstates $\zeta_{n_1 ,n_2}(0,0;\vec{q},-\vec{q})$ is given by:
\begin{equation}
	\label{Gamma_s}
	\zeta_{n_1 ,n_2}(0,0;\boldsymbol{q},-\boldsymbol{q})=\int d\boldsymbol{r} 
	\times \Psi^*_{n_1,\boldsymbol{q}}(\boldsymbol{r}) \Psi^*_{n_2,-\boldsymbol{q}}(\boldsymbol{r}) \Psi_{d,0}(\boldsymbol{r}) \Psi_{d,0}(\boldsymbol{r})
\end{equation}
where $\Psi_{n,\boldsymbol{q}}(\boldsymbol{r})$ is Bloch function of the eigenstate at quasi-momentum $\boldsymbol{q}$ in the $n$ band. In the calculation, we assume the periodic boundary conditions, and consider that 
$|\zeta_{n_1 ,n_2}(0,0;\boldsymbol{q},-\boldsymbol{q})|^2=\int d\boldsymbol{r}\times |\Psi^*_{n_1,\boldsymbol{q}}(\boldsymbol{r}) \Psi^*_{n_2,-\boldsymbol{q}}(\boldsymbol{r}) \Psi_{d,0}(\boldsymbol{r}) \Psi_{d,0}(\boldsymbol{r})|^2$. 
After neglecting the scattering channels of higher bands, we calculate the scattering channels, as shown in figure \ref{fig:scatter}CD. In the square lattice, the cross-section of the strongest channel $SS$, $P_1P_1$ and $P_2P_2$ are all around $10\%$ of the total cross-section respectively. Besides, there are many smaller channels included in 'Others.' There is no significant difference in scattering cross-section values among the first six channels, which means no dominant scattering channel in a square lattice.
By contrast, in the triangular lattice, the proportion of scattering cross-section of the $SS$ channel is $38.5\%$, while that of the second strong channel $D_1S$ is only $9.8\%$.
Besides, the proportion of other channels is much lower than that of channel $SS$. Consequently, the channel $SS$ is dominant in the two-body scattering process of the triangular optical lattice. The experimental results are consistent with the theoretical results, as shown in figure \ref{fig:scatter}F. For example, at time $t_{evo}=12$ ms, the experimental proportion of the S-band is 55.8\%, which is roughly equal to the theoretical prediction of 57.3\%\cite{Guo2021}.

\subsection{Quantum dynamical oscillations of ultracold atoms in the F and D bands}
Here we review the observation of quantum dynamical oscillations of ultracold atoms in the F and D bands of the 1D optical lattice\cite{PhysRevA.94.033624}. We can control the Bragg reflections at the Brillouin-zone edge up to the third order and observe three different types of quantum oscillations\cite{PhysRevA.94.033624}.

\begin{figure}[h!]
	\begin{center}
		\includegraphics[width=14 cm]{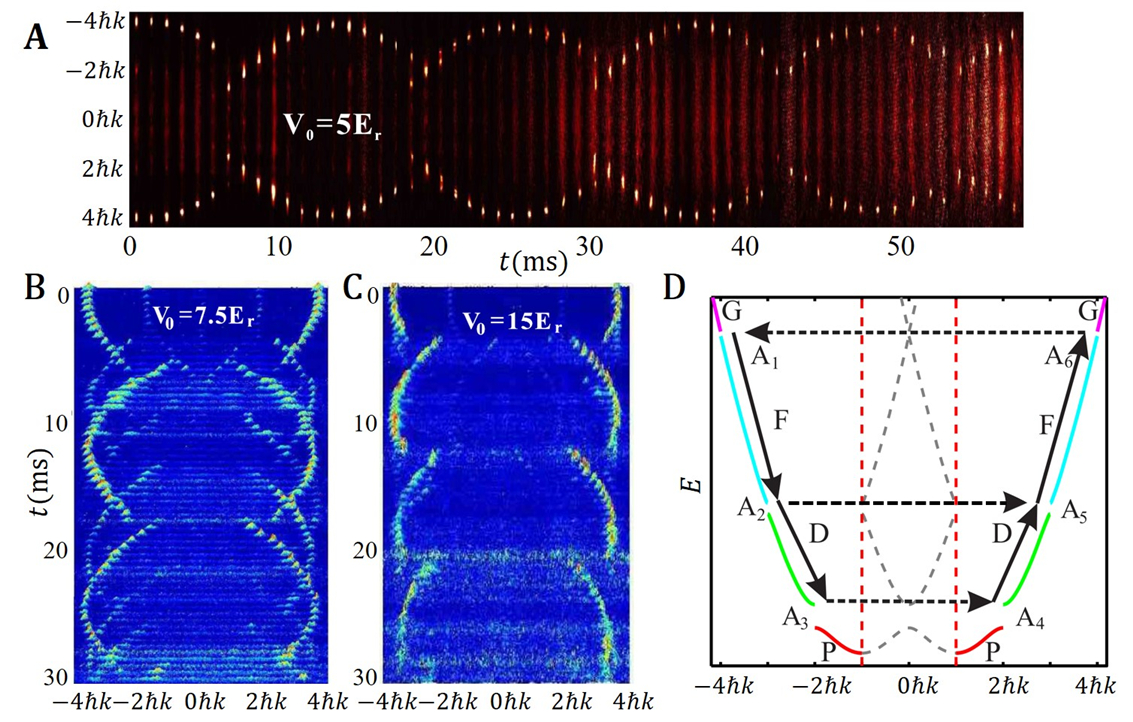}
	\end{center}
	\caption{Quantum oscillations of the BECs in higher bands of the 1D optical lattice. Experimental results in momentum space with lattice depth (\textbf{A}) $5E_r$, (\textbf{B}) $7.5E_r$, and (\textbf{C}) $15E_r$. (\textbf{D}) Schematic of extended Bloch bands of the 1D optical lattice.}
	\label{fig:FDoscillation}
\end{figure}

The BECs is initially loaded in the G band, where the atoms mostly populate at momenta $\rm4\hbar k$, as shown in figure \ref{fig:shortcut_pulse}B(2). Then the atoms fall into the F-band due to the small gap between G- and F-bands. The following trace of the atoms is shown in the extended band structures as shown in figure \ref{fig:FDoscillation}D. Once the BECs are in the F-band, it loses momentum while gaining potential energy from the harmonic confinement ($A_1\to A_2$). Once arriving $A_2$, the atoms face different dynamics depending on the lattice depth. When the depth is small ($\sim5E_r$), and the Bragg reflections at $A_2$ are weak, the BECs will continue into the D-band by a Landau-Zener transition and reach $A_3$. Then the atoms will be Bragg reflected to $A_4$ and reverse its dynamics ($A_4\to A_5 \to A_6$). This oscillation is shown in figure \ref{fig:FDoscillation}A. When the lattice is strong ($\sim15E_r$), and the gap at $A_2$ is large, the Bragg reflection can dominate the dynamics, forbidding the atoms from tunneling from the F band to the D band. Instead, the atoms at $A_2$ will transfer to $A_5$ and oscillate only within the F-band, which is shown in figure \ref{fig:FDoscillation}C. For intermediate depth  ($\sim7.5E_r$), these two oscillation modes exist simultaneously, as shown in figure \ref{fig:FDoscillation}B\cite{PhysRevA.94.033624}.

\subsection{Nonlinear dynamical evolution for P-band ultracold atoms in 1D optical lattice}
\begin{figure}[h!]
	\begin{center}
		\includegraphics[width=14 cm]{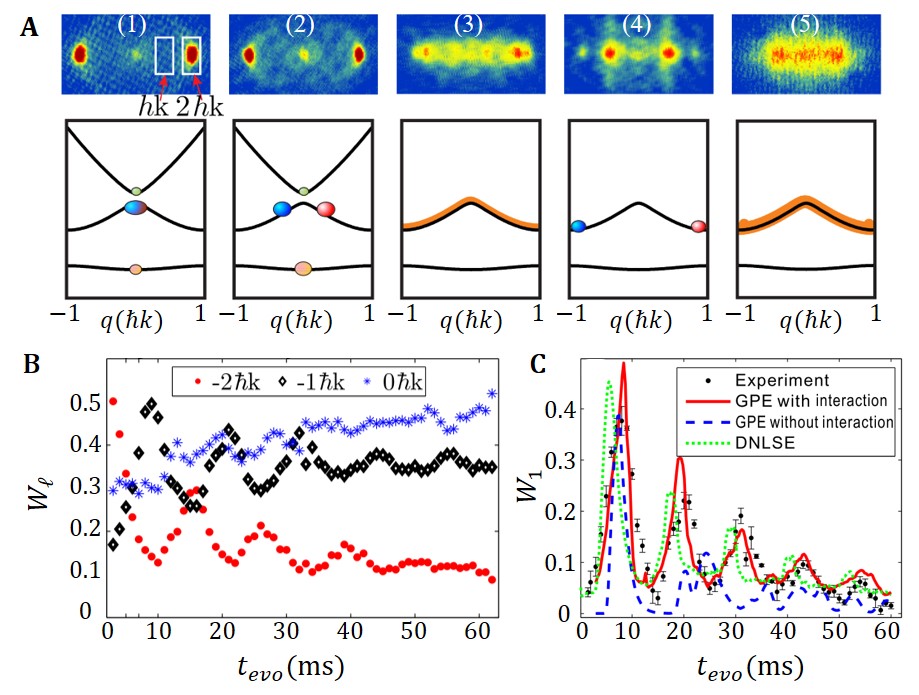}
	\end{center}
	\caption{(\textbf{A}) First row: the measured momentum distributions of the BEC in the P-band of the 1D lattice at different holding times (0 ms, 2 ms, 5 ms, 7 ms, and 30 ms), and the white rectangles are the region for us to calculate the proportion of different momentum states; the second row: schematic illustration of the corresponding population distributions in the Bloch band. (\textbf{B}) Population oscillations around momenta $0\hbar k$ (blue stars), $-\hbar k$ (black diamonds), and $-2\hbar k$ (red dots) with $t > 2$ ms. The oscillation period is about 14.9 ms with $V = 5E_r$. (\textbf{C}) Population for $\hbar k$ vs time $t_{evo}$. The experimental results and theoretical simulations with and without the interaction correspond to the black dots with an error bar, red solid curves, and blue dashed curves, respectively.}
	\label{fig:p_evolution}
\end{figure}

The dynamical evolution for atoms in P-band is different from that of F- and D-bands\cite{Hu2015}. After loading the BECs into the P-band with zero quasi-momentum for the lattice depth $5E_r$, we hold the BECs for time $t_{evo}$ and then measure the momentum distribution. The momentum distributions at different holding time $t_{evo}$ are shown in figure \ref{fig:p_evolution}A. We use the normalized populations $W_\ell(t_{evo})$ of momentum states $| \ell\cdot 2 \hbar k + q \rangle$ to better quantify the dynamical evolution of atoms. At the beginning of the evolution process $t_{evo}<1.5$ ms, we find $W_\ell(t_{evo})$ oscillating rapidly, as shown in figure \ref{fig:symmetry}C and figure \ref{fig:p_evolution}A(1)(2). Then, after a short transition time, a different type of oscillation begins to emerge around $t_{evo}=2$ ms.The period of this oscillation is 14.9 ms, which is shown in figure \ref{fig:p_evolution}A(3-5) and \ref{fig:p_evolution}B.

The short-period oscillations shown in figure \ref{fig:symmetry}C are beating signals due to the coherent superposition of different bands. From the numerical analysis, the superposed state is close to $|\psi(t_{evo}=0)=\sqrt{0.9}\Psi_{2,0}+\sqrt{0.05}\Psi_{1,0}+\sqrt{0.05}\Psi_{3,0}$. The rapid oscillations disappear at about 1.5 ms. Then a long-period oscillation begins to emerge at around 2 ms. There are five cycles of the long-period oscillation in figure \ref{fig:p_evolution}C. The long-period oscillation reflects the random phase between neighboring lattice sites, which can be well captured by the simulation with the Gross-Pitaevskii equation (shown by the red line in figure \ref{fig:p_evolution}C). This experiment paves the way to study the long-time dynamical evolution of the high orbital physics for other novel quantum states, such as the sliding phase\cite{Hu2015}.

\section{Discussion and Conclusions}\label{summary}
In this review, we concentrate on the methods to prepare and control the BEC in optical lattices linked to one-body physics. The many-body interactions also play an important role in the system of ultracold atoms in the optical lattices. For the control methods mentioned in this review, such as the shortcut method, amplitude modulation, and nonadiabatic holonomic control, the operation time ($<1$ ms) is much shorter than the time when the interaction has a significant effect. Therefore, the control schemes designed ignoring the influence of many-body interaction is still very successful. The effects of the interaction mainly occur in the long-term evolution process in the optical lattice after we manipulate or prepare the Bloch states, such as the de-coherence in the Ramsey interferometry with motional states, the two-body collision of D-band atoms, and the emergency of the exotic quantum states of p-orbitals. On the other hand, we can also utilize the interactions to expand the methods of atomic manipulation in optical lattices. For example, a two-qubit gate can be achieved by adapting the interaction scheme based on the method shown in section \ref{sec:gate}. Considering two nearby sites, denoted as $a$ and $b$, the orbital states are $|s_a s_b\rangle$, $|d_a d_b\rangle$, $|d_a s_b\rangle$, $|s_a d_b\rangle$. The relevant interactions between neighboring sites contain this term $U_0(s^\dagger_a d^\dagger_b d_b s_a+d^\dagger_a s^\dagger_b s_b d_a)$. A $\sqrt{\rm{swap}}$ gate control can be reached by letting atoms interact for a time duration $\pi\hbar/(4U_0)$. In a word, we can use these control methods to observe the special dynamic mechanism and novel quantum states produced by the interaction of different orbitals of optical lattices, and we can also use the interactions to achieve more manipulation.

In summary, we review our practical methods for manipulating the high orbitals of ultracold atoms in optical lattices. The shortcut method is characterized by short time and high fidelity, which can directly transfer ultracold atoms from the ground state in the harmonic trap to any Bloch state, and accurately manipulate atoms of different orbitals in optical lattices. This method can be used to construct atomic orbital interferometers and qubits and to study the dynamic properties of high orbital atoms in optical lattices. The band swapping technique considers the interaction between atoms and the additional potential trap (such as harmonic trap) besides the optical lattice, which is more suitable for studying the ground and metastable states of the system. The amplitude modulation focuses on coupling different Bloch bands and can be used to realize quantum gates and the large-momentum-transfer beam splitter. Based on these methods, the atom-orbital qubit under nonadiabatic holonomic quantum control and Ramsey interferometry with trapped motional quantum states of the optical lattice can be constructed. Many exotic quantum states of the high orbital atoms have been observed. Then we study the quantum dynamical evolution of atoms in high bands. The effective manipulation of the high orbitals provides strong support for applying the ultracold atoms in the optical lattice in quantum simulation, quantum computing, and quantum precision measurement.

\section*{Funding}
This work is supported by the National Natural Science Foundation of China (Grants No. 12104020, No. 61727819, and No. 11334001), the National Basic Research Program of China (Grants No. 2021YFA0718300 and No. 2021YFA1400901), the Project funded by China Postdoctoral Science Foundation (Grant No. 2020TQ0017), the Science and Technology Major Project of Shanxi (No. 202101030201022), and the Space Application System of China Manned Space Program.

\section*{Acknowledgments}
Thank Prof. Yiqiu Wang for establishing this experimental group of ultra-cold atoms in China. In completing the above work, thank Xiaopeng Li, J\"org Schmiedmayer, Biao Wu, Hongwei Xiong, Guangjiong Dong, Peng Zhang, and Lan Yin.
\newpage
\section*{References}
\bibliographystyle{unsrt}
\bibliography{test}

\begin{thebibliography}{10}

\bibitem{Greiner2002}
Markus Greiner, Olaf Mandel, Tilman Esslinger, Theodor~W. H{\"a}nsch, and
  Immanuel Bloch.
\newblock Quantum phase transition from a superfluid to a mott insulator in a
  gas of ultracold atoms.
\newblock {\em Nature}, 415(6867):39--44, Jan 2002.

\bibitem{Bloch2012}
Immanuel Bloch, Jean Dalibard, and Sylvain Nascimb{\`e}ne.
\newblock Quantum simulations with ultracold quantum gases.
\newblock {\em Nature Physics}, 8(4):267--276, Apr 2012.

\bibitem{RevModPhys.80.885}
Immanuel Bloch, Jean Dalibard, and Wilhelm Zwerger.
\newblock Many-body physics with ultracold gases.
\newblock {\em Rev. Mod. Phys.}, 80:885--964, Jul 2008.

\bibitem{Goldman2016}
N.~Goldman, J.~C. Budich, and P.~Zoller.
\newblock Topological quantum matter with ultracold gases in optical lattices.
\newblock {\em Nature Physics}, 12(7):639--645, Jul 2016.

\bibitem{RevModPhys.86.153}
I.~M. Georgescu, S.~Ashhab, and Franco Nori.
\newblock Quantum simulation.
\newblock {\em Rev. Mod. Phys.}, 86:153--185, Mar 2014.

\bibitem{doi:10.1126/science.aal3837}
Christian Gross and Immanuel Bloch.
\newblock Quantum simulations with ultracold atoms in optical lattices.
\newblock {\em Science}, 357(6355):995--1001, 2017.

\bibitem{PhysRevLett.124.120403}
E.~R. Moan, R.~A. Horne, T.~Arpornthip, Z.~Luo, A.~J. Fallon, S.~J. Berl, and
  C.~A. Sackett.
\newblock Quantum rotation sensing with dual sagnac interferometers in an
  atom-optical waveguide.
\newblock {\em Phys. Rev. Lett.}, 124:120403, Mar 2020.

\bibitem{Shui2021}
Hongmian Shui, Shengjie Jin, Zhihan Li, Fansu Wei, Xuzong Chen, Xiaopeng Li,
  and Xiaoji Zhou.
\newblock Atom-orbital qubit under nonadiabatic holonomic quantum control.
\newblock {\em Phys. Rev. A}, 104:L060601, Dec 2021.

\bibitem{2011_Wirth_NatPhys}
G.~Wirth, M.~Ölschläger, and A.~Hemmerich.
\newblock Evidence for orbital superfluidity in the p-band of a bipartite
  optical square lattice.
\newblock {\em Nat. Phys.}, 7(2):147--153, 2011.

\bibitem{Niu_PRL2018}
Linxiao Niu, Shengjie Jin, Xuzong Chen, Xiaopeng Li, and Xiaoji Zhou.
\newblock Observation of a dynamical sliding phase superfluid with $p$-band
  bosons.
\newblock {\em Phys. Rev. Lett.}, 121:265301, Dec 2018.

\bibitem{Wang2021}
Xiao-Qiong Wang, Guang-Quan Luo, Jin-Yu Liu, W.~Vincent Liu, Andreas Hemmerich,
  and Zhi-Fang Xu.
\newblock Evidence for an atomic chiral superfluid with topological
  excitations.
\newblock {\em Nature}, 596(7871):227--231, Aug 2021.

\bibitem{Jin_PRL2021}
S.~Jin, W.~Zhang, X.~Guo, X.~Chen, X.~Zhou, and X.~Li.
\newblock Evidence of potts-nematic superfluidity in a hexagonal $s{p}^{2}$
  optical lattice.
\newblock {\em Phys. Rev. Lett.}, 126:035301, Jan 2021.

\bibitem{Hu2018}
Dong Hu, Linxiao Niu, Shengjie Jin, Xuzong Chen, Guangjiong Dong, J\"{o}rg
  Schmiedmayer, and Xiaoji Zhou.
\newblock Ramsey interferometry with trapped motional quantum states.
\newblock {\em Communications Physics}, 1(1), June 2018.

\bibitem{NJP_zhou}
X.~Zhou, S.~Jin, and J.~Schmiedmayer.
\newblock Shortcut loading a {Bose-Einstein} condensate into an optical
  lattice.
\newblock {\em New J. Phys.}, 20(5):055005, 2018.

\bibitem{PhysRevLett.99.200405}
Torben M\"uller, Simon F\"olling, Artur Widera, and Immanuel Bloch.
\newblock State preparation and dynamics of ultracold atoms in higher lattice
  orbitals.
\newblock {\em Phys. Rev. Lett.}, 99:200405, Nov 2007.

\bibitem{LXX2011}
Xinxing Liu, Xiaoji Zhou, Wei Xiong, Thibault Vogt, and Xuzong Chen.
\newblock Rapid nonadiabatic loading in an optical lattice.
\newblock {\em Phys. Rev. A}, 83:063402, Jun 2011.

\bibitem{PhysRevA.87.063638}
Yueyang Zhai, Xuguang Yue, Yanjiang Wu, Xuzong Chen, Peng Zhang, and Xiaoji
  Zhou.
\newblock Effective preparation and collisional decay of atomic condensates in
  excited bands of an optical lattice.
\newblock {\em Phys. Rev. A}, 87:063638, Jun 2013.

\bibitem{Hu2015}
Dong Hu, Linxiao Niu, Baoguo Yang, Xuzong Chen, Biao Wu, Hongwei Xiong, and
  Xiaoji Zhou.
\newblock Long-time nonlinear dynamical evolution for $p$-band ultracold atoms
  in an optical lattice.
\newblock {\em Phys. Rev. A}, 92:043614, Oct 2015.

\bibitem{Yang2016}
Baoguo Yang, Shengjie Jin, Xiangyu Dong, Zhe Liu, Lan Yin, and Xiaoji Zhou.
\newblock Atomic momentum patterns with narrower intervals.
\newblock {\em Phys. Rev. A}, 94:043607, Oct 2016.

\bibitem{PhysRevA.94.033624}
Zhongkai Wang, Baoguo Yang, Dong Hu, Xuzong Chen, Hongwei Xiong, Biao Wu, and
  Xiaoji Zhou.
\newblock Observation of quantum dynamical oscillations of ultracold atoms in
  the $f$ and $d$ bands of an optical lattice.
\newblock {\em Phys. Rev. A}, 94:033624, Sep 2016.

\bibitem{Guo2021}
Xinxin Guo, Zhongcheng Yu, Peng Peng, Guoling Yin, Shengjie Jin, Xuzong Chen,
  and Xiaoji Zhou.
\newblock Dominant scattering channel induced by two-body collision of $d$-band
  atoms in a triangular optical lattice.
\newblock {\em Phys. Rev. A}, 104:033326, Sep 2021.

\bibitem{doi:10.1126/sciadv.1500854}
Shintaro Taie, Hideki Ozawa, Tomohiro Ichinose, Takuei Nishio, Shuta Nakajima,
  and Yoshiro Takahashi.
\newblock Coherent driving and freezing of bosonic matter wave in an optical
  lieb lattice.
\newblock {\em Science Advances}, 1(10):e1500854, 2015.

\bibitem{PhysRevA.72.053605}
A.~Browaeys, H.~H\"affner, C.~McKenzie, S.~L. Rolston, K.~Helmerson, and W.~D.
  Phillips.
\newblock Transport of atoms in a quantum conveyor belt.
\newblock {\em Phys. Rev. A}, 72:053605, Nov 2005.

\bibitem{PhysRevLett.106.015302}
Matthias \"Olschl\"ager, Georg Wirth, and Andreas Hemmerich.
\newblock Unconventional superfluid order in the $f$ band of a bipartite
  optical square lattice.
\newblock {\em Phys. Rev. Lett.}, 106:015302, Jan 2011.

\bibitem{Niu2015}
Linxiao Niu, Dong Hu, Shengjie Jin, Xiangyu Dong, Xuzong Chen, and Xiaoji Zhou.
\newblock Excitation of atoms in an optical lattice driven by polychromatic
  amplitude modulation.
\newblock {\em Opt. Express}, 23(8):10064--10074, Apr 2015.

\bibitem{PhysRevLett.94.080403}
Michael K\"ohl, Henning Moritz, Thilo St\"oferle, Kenneth G\"unter, and Tilman
  Esslinger.
\newblock Fermionic atoms in a three dimensional optical lattice: Observing
  fermi surfaces, dynamics, and interactions.
\newblock {\em Phys. Rev. Lett.}, 94:080403, Mar 2005.

\bibitem{PhysRevLett.74.1542}
A.~Kastberg, W.~D. Phillips, S.~L. Rolston, R.~J.~C. Spreeuw, and P.~S. Jessen.
\newblock Adiabatic cooling of cesium to 700 nk in an optical lattice.
\newblock {\em Phys. Rev. Lett.}, 74:1542--1545, Feb 1995.

\bibitem{PhysRevLett.87.160405}
Markus Greiner, Immanuel Bloch, Olaf Mandel, Theodor~W. H\"ansch, and Tilman
  Esslinger.
\newblock Exploring phase coherence in a 2d lattice of bose-einstein
  condensates.
\newblock {\em Phys. Rev. Lett.}, 87:160405, Oct 2001.

\bibitem{Robert_de_Saint_Vincent_2010}
M.~Robert de~Saint-Vincent, J.-P. Brantut, Ch.~J. Bord{\'{e}}, A.~Aspect,
  T.~Bourdel, and P.~Bouyer.
\newblock A quantum trampoline for ultra-cold atoms.
\newblock {\em {EPL} (Europhysics Letters)}, 89(1):10002, jan 2010.

\bibitem{Impens_2011}
Fran{\c{c}}ois Impens, Franck Pereira~Dos Santos, and Christian~J Bord{\'{e}}.
\newblock The theory of quantum levitators.
\newblock {\em New Journal of Physics}, 13(6):065024, jun 2011.

\bibitem{RevModPhys.81.1051}
Alexander~D. Cronin, J\"org Schmiedmayer, and David~E. Pritchard.
\newblock Optics and interferometry with atoms and molecules.
\newblock {\em Rev. Mod. Phys.}, 81:1051--1129, Jul 2009.

\bibitem{PhysRevLett.67.177}
F.~Riehle, Th. Kisters, A.~Witte, J.~Helmcke, and Ch.~J. Bord\'e.
\newblock Optical ramsey spectroscopy in a rotating frame: Sagnac effect in a
  matter-wave interferometer.
\newblock {\em Phys. Rev. Lett.}, 67:177--180, Jul 1991.

\bibitem{PhysRevLett.83.2745}
C.~S. O'Hern, T.~C. Lubensky, and J.~Toner.
\newblock Sliding phases in $\mathit{XY}$ models, crystals, and cationic
  lipid-dna complexes.
\newblock {\em Phys. Rev. Lett.}, 83:2745--2748, Oct 1999.

\bibitem{PhysRevB.33.4767}
Enzo Granato and J.~M. Kosterlitz.
\newblock Critical behavior of coupled xy models.
\newblock {\em Phys. Rev. B}, 33:4767--4776, Apr 1986.

\bibitem{PhysRevB.31.4516}
M.~Y. Choi and S.~Doniach.
\newblock Phase transitions in uniformly frustrated xy models.
\newblock {\em Phys. Rev. B}, 31:4516--4526, Apr 1985.

\bibitem{FEIGELMAN1990177}
M.V. Feigel'man, V.B. Geshkenbein, and A.I. Larkin.
\newblock Pinning and creep in layered superconductors.
\newblock {\em Physica C: Superconductivity}, 167(1):177--187, 1990.

\bibitem{PhysRevB.12.877}
R.~A. Klemm, A.~Luther, and M.~R. Beasley.
\newblock Theory of the upper critical field in layered superconductors.
\newblock {\em Phys. Rev. B}, 12:877--891, Aug 1975.

\bibitem{PhysRevLett.73.1384}
T.~Stoebe, P.~Mach, and C.~C. Huang.
\newblock Unusual layer-thinning transition observed near the
  smectic-$a$-isotropic transition in free-standing liquid-crystal films.
\newblock {\em Phys. Rev. Lett.}, 73:1384--1387, Sep 1994.

\bibitem{PhysRevLett.59.1112}
Ming Cheng, John~T. Ho, S.~W. Hui, and Ronald Pindak.
\newblock Electron-diffraction study of free-standing liquid-crystal films.
\newblock {\em Phys. Rev. Lett.}, 59:1112--1115, Sep 1987.

\bibitem{PhysRevLett.80.4345}
C.~S. O'Hern and T.~C. Lubensky.
\newblock Sliding columnar phase of dna-lipid complexes.
\newblock {\em Phys. Rev. Lett.}, 80:4345--4348, May 1998.

\end{thebibliography}

\newpage
\section*{Figure captions}

\end{document}